\documentclass[twocolumn]{aa} 
\usepackage{upgreek}
\usepackage{graphicx}
\usepackage{placeins}
\usepackage{txfonts}
\usepackage{multirow} 
\usepackage{textcomp} 
\usepackage[colorlinks=true,allcolors=blue,linkcolor=red,urlcolor=blue]{hyperref}
\usepackage{natbib}
\usepackage{ulem}
\usepackage{booktabs}
\usepackage{amsmath}
\usepackage{xcolor}
\usepackage{amssymb}
\usepackage{bm}
\usepackage{txfonts}
\usepackage{longtable}
\usepackage{microtype}
\usepackage{lipsum}
\usepackage{subcaption}   
\usepackage{lscape}
\usepackage{placeins} 

\makeatletter
\renewcommand\sout{\bgroup\markoverwith{\textcolor{red}{\rule[0.5ex]{2pt}{0.8pt}}}\ULon}
\makeatother

\newcommand{\si}{\ion{S}{I}~}
\newcommand{\ab}{ab initio~}
\newcommand{\ft}{fine-tuned~}

\begin{document}

\title{Theoretical investigation of transition data of
astrophysical importance in neutral sulphur}

\author{W. Li\inst{1} \and A. M. Amarsi\inst{2} \and P. J\"onsson\inst{3}}

\institute{
$^{1}$State Key Laboratory of Solar Activity and Space Weather, National Astronomical Observatories, Chinese Academy of Sciences \\
$^{2}$Theoretical Astrophysics, Department of Physics and Astronomy, Uppsala
University, Box 516, SE-751 20 Uppsala, Sweden \\
$^{3}$Department of Materials Science and Applied Mathematics, Malm\"o University, SE-205 06 Malm\"o, Sweden\\
}

\abstract
{Accurate and comprehensive atomic data are essential for 
the modelling of stellar spectra.
Uncertainties in the oscillator strengths of specific lines used for abundance analyses directly
translate into uncertainties in the derived elemental 
abundances; incomplete or biased atomic data
sets can impart significant errors in non-local thermodynamic equilibrium (non-LTE)
modelling.
Theoretical calculations of atomic data are therefore crucial to 
supplement the limited experimental results.
In this work, we present extensive
atomic data, including oscillator strengths, transition
rates, and lifetimes for 1,730 electric-dipole (E1) transitions among 107
levels in neutral sulphur (\ion{S}{I}) using the multi-configuration
Dirac–Hartree–Fock (MCDHF) and relativistic-configuration-interaction (RCI)
methods. These levels belong to the configurations $\mathrm{3p^3np~(n=3-7)}$,
$\mathrm{3p^3nf~(n=4,5)}$, $\mathrm{3s3p^5}$, $\mathrm{3p^3ns~(n=4-7)}$, and
$\mathrm{3p^3nd~(n=3-6)}$. 
The accuracy of the computed
transition rates is assessed by combining the comparison of the differences in
transition rates between the Babushkin and Coulomb gauges with a cancellation-factor (CF) analysis. 
Approximately 16\% of the ab initio results achieved an accuracy classification of A-B, 
corresponding to uncertainties within 10\%, as defined by the Atomic Spectra Database 
of the National Institute of Standards and Technology (NIST ASD).
Applying a fine-tuning technique was found
to significantly improve the 
accuracy of the results in the Coulomb gauge, thereby improving the consistency 
between the Babushkin and Coulomb gauges;
about 24\% of the fine-tuned transition data are assigned to the accuracy classes A-B. 
}

\keywords{Atomic data --- Atomic processes --- Methods: numerical --- Sun: abundances --- Stars: abundances}

\date{Received 2 September 2025 / Accepted 26 January 2026}
\maketitle

\section{Introduction}\label{sec:intro}
Spectroscopy is a powerful tool in astrophysics, providing critical insights 
into the physical conditions and chemical abundances of stars \citep{2019asia.book.....T}. 
Sulphur, as one of the most abundant elements in the Universe, exhibits a wealth of spectral 
features across multiple wavelength bands that have been observed with both 
space- and ground-based facilities \citep[e.g.][]{1983ApJ...267L.125D, 1993ApJ...411..655S,1994ApJ...428..329C}. 
These lines serve as powerful diagnostics for probing stellar atmospheres, the 
interstellar medium, and galaxies \citep[e.g.][]{1988MNRAS.231..419J, 2002ApJ...570..439F, Anderson_2013}. 
In particular, 
as an $\upalpha$ element,
sulphur is routinely employed in the study of
Galactic chemical evolution 
\citep[e.g.][]{2007A&A...469..319N,
2015A&A...580A.129S,2017A&A...604A.128D,
2020A&A...634A.136C,2023A&A...678A.195D}.

Given this importance, it is of interest to reliably infer
sulphur abundances in stars and in other astronomical objects.
However, accurate analyses of observed sulphur spectra require accurate
and complete atomic data sets. Errors in transition data of small sets of diagnostic lines
directly impact inferred sulphur abundances;
while incomplete or systematically biased atomic
data sets could impact 
the statistical equilibrium
in non-local thermodynamic equilibrium (non-LTE) models
\citep[e.g.][]{2001A&A...375..899K,2003A&A...399..525D,
2005PASJ...57..751T,2024ARep...68.1159K}.
Considerable effort has therefore been
devoted to both theoretical calculations and 
laboratory measurements of transition data for neutral sulphur.

On the experimental side, earlier measurements of oscillator strength and transition 
probabilities between multiplet transitions were reported by 
\cite{1966ApJ...146..940S},
\cite{1967PhRv..159...31B}, and \cite{1968ZNatA..23.1707M}.
\cite{1990JGR....9521313D} measured relative oscillator strengths for the 
$\mathrm{3p^4\,^3P - 3p^3(^4S)4s\,^3S^o}$\,(1814 \,\text{\AA}), 
$\mathrm{3p^4\,^3P - 3p^3(^2D)4s\,^3D^o}$\,(1479 \,\text{\AA}), 
and $\mathrm{3p^4\,^3P - 3p^3(^4S)3d \, ^3D^o}$ (1429 \,\text{\AA}) multiplets 
using the method of electron energy-loss spectroscopy. \cite{1990PhyS...42..540D} 
determined the radiative lifetimes of eight different multiplets in the visible range 
using the high-frequency deflection technique.
\cite{1994ApJ...428..393B} derived oscillator strengths for eight transitions of the
multiplets between $\mathrm{3p^{3}(^{4}S)4s~^{3}S^o}$, 
$\mathrm{3p^{3}(^{2}P)4s~^{3}P^o}$, and $\mathrm{3p^{4}~^3P}$
based on measured mean lifetimes and branching ratios using beam-foil spectroscopic 
techniques at the Toledo Heavy Ion Accelerator. 
Based on the oscillator strength reported  by \cite{1990JGR....9521313D} and \cite{1994ApJ...428..393B}, 
\cite{1995ApJ...452..269F} analysed interstellar spectra 
towards $\zeta$-Oph acquired with the Goddard High-Resolution Spectrograph and obtained
a self-consistent set of oscillator strengths for approximately a dozen \ion{S}{I}~ lines from a
curve of growth. 
More recently, using time-resolved vacuum-ultraviolet laser spectroscopy, 
\cite{RZerne_1997} determined oscillator strengths for the 
$\mathrm{3p^3(^4S)4s~^3S^o_1 - 3p^3(^4S)4p~^3P_{0,1,2}}$ triplet at 10450 \AA~and 
\cite{1997PhRvA..55.1836B} reported lifetimes for twelve states. These lifetime data 
were combined by \cite{1998ApJ...502.1010B} with their theoretical branching ratios 
from relativistic Hartree--Fock (HFR) calculations to deduce a consistent set of 
oscillator strengths for vacuum ultraviolet lines.

Besides experimental measurements, there are also a number of theoretical calculations
of the transition data for \ion{S}{I}. 
\cite{1982PhLA...87..394G} calculated the optical oscillator strengths for the sulphur 
isoelectric sequence using an analytic atomic independent-particle-model potential. 
\cite{1975SAOSR.362.....K} calculated the oscillator strength semi-empirically 
using scaled Thomas-Fermi-Dirac radial wave functions and eigenvectors found 
through least-squares fits to observed energy levels.
\cite{1985ApJ...290..424H},
\cite{1999PhST...83...44F},
\cite{1998ApJ...497..493T}, and 
\cite{1997JPhB...30.3873C} did the calculations using the 
CIV3 code based on configuration–interaction (CI) method \citep{HIBBERT1975141}.
\cite{1986ADNDT..35..185F} calculated the oscillator strengths for allowed n=3 -- 3 and 
3 -- 4 transitions of \si with the HFR program package.
\cite{Fischer_1987} carried out multi-configuration Hartree–Fock (MCHF) calculations 
for a number of $\mathrm{3p^3(^2D)nd~^3P^o}$ terms in \si. 
More recently, \cite{FROESEFISCHER2006607} reported both allowed electric-dipole (E1) 
and some forbidden transitions among levels up to $\mathrm{3p^3(^4S)3d~^3D^o}$ by 
including relativistic effects through the Breit-Pauli Hamiltonian.
\cite{2006JPhB...39.2861Z}, hereafter ZB2006 reported fine structure 
energies for configurations $\mathrm{3p^3nl}$ of certain levels up to n=12 and 
the associated E1 oscillator strengths among them using a B-spline box-based multi-channel method. 
In this calculation, they included CI effects by using a fairly large B-spline basis 
set and relativistic effects through the Breit–Pauli Hamiltonian. 
\cite{2006JPhB...39.4301D} and \cite{2008ADNDT..94..561D}, hereafter DH2008
presented oscillator strengths and radiative rates for 2173 E1 transitions among 120 
levels using the CIV3 program with a fine-tuning technique. 
However, as reported by ZB2006, comparisons of oscillator strengths and transition 
probabilities among various calculations sometimes show large disagreement.
The large disagreement between different calculations indicates large uncertainties in the 
theoretical atomic data, which may have a significant impact on astrophysical spectroscopic analyses.
This may be due to the insufficient configuration interactions included in many of the 
calculations. The most recent work by DH2008 included 300,000 configuration state functions (CSFs) in their CI 
calculation, which, to our knowledge, is the largest scale calculation for \si to date.
However, larger scale calculations, such as those with millions of CSFs, had not been performed until now.

The purpose of this study is to present the results of our extensive calculations 
using the fully relativistic multi-configuration Dirac–Hartree–Fock (MCDHF) and 
relativistic configuration interaction (RCI) methods. 
Here we provide two sets of calculation results. One set was obtained directly from 
the MCDHF and RCI calculations, which we refer to as the ab initio results. 
The other set involves fine-tuning of eigenvalues in the RCI calculations, 
which we refer to as the fine-tuned results. We compare the fine-tuned data with the 
ab initio results to investigate the effects of fine-tuning on radiative data and lifetimes.
The new atomic data set may help with assessing the accuracy of various atomic data via
comparisons with other data sets.

This paper is structured into four sections, including the Introduction. 
Section \ref{sec:theory} introduces the theoretical methods and computational schemes. 
Section \ref{sec:results} presents the results and discussions on energy levels, transition data, 
and lifetimes. Section \ref{sec:conclusion} includes our conclusion.

\begin{table*}
    \caption{Summary of the computational schemes for \si{}.}
    \centering
    \begin{tabular}{ccccccccccccc}
    \hline\midrule
   Parity & MR-MCDHF   & MR-RCI & AS & $N_\mathrm{{CSFs}}$ \\
\midrule
       &    $\mathrm{3s^23p^3np~(n=3-8)}$,     & $\mathrm{3p^6}$, ${\mathrm{3s^23p^3np~(n=3-8)}}$, $\mathrm{3s^23p^3nf~(n=4,6)}$,&                              &               \\
even   &    $\mathrm{3s^23p^3nf~(n=4,5)}$,  &  $\mathrm{3s3p^3np7d~(n=3-7)}$, $\mathrm{3sp^3nf7d~(n=4,5)}$,   &   \{$\mathrm{12s, 12p, 11d, 10f, 9g, 8h}$\} &  11 049 420   \\
       &     $\mathrm{3p^6}$                    &       $\mathrm{3s^23pnp7d^2~(n=3,4)}$  &                                     &     \\
\midrule
       &   $\mathrm{3s^23p^3ns~(n=4-7)}$, & $\mathrm{3s3p^5}$, $\mathrm{3s^23p^3ns~(n=4-8)}$, &                              &               \\
odd    &   $\mathrm{3s^23p^3nd~(n=3-7)}$  & $\mathrm{3p^54s}$, $\mathrm{3s^23p^3nd~(n=3-7)}$, &   \{$\mathrm{12s, 12p, 11d, 10f, 9g, 8h}$\}  &  8 146 099    \\
       &   $\mathrm{3s3p^5}$, $\mathrm{3p^54s}$    & $\mathrm{3s3p^3ns7d~(n=4-7)}$, $\mathrm{{3sp^3nd7d~(n=3-7)}}$, &                         &               \\
       &                                       & $\mathrm{3p^57d}$, $\mathrm{3s^23p4s7d^2}$, $\mathrm{3s3p^4np~(n=4-8)}$   &                         &               \\

\bottomrule
    \end{tabular}
    \label{tab:MR}
    \tablefoot{MR-MCDHF and MR-RCI, respectively, denote the multi-reference sets used in the MCDHF and RCI calculations. 
    AS is the active sets of orbitals. $N_\mathrm{CSFs}$ are the numbers of generated CSFs in the final RCI calculations.}
\end{table*}

\section{Method}\label{sec:theory}

\subsection{Multi-configuration Dirac–Hartree–Fock approach}\label{sec:MCDHF}
The calculations were performed using the MCDHF and RCI methods, based on a Dirac–Coulomb 
Hamiltonian, which are implemented in the general-purpose relativistic atomic-structure 
package \textsc{Grasp2018}\footnote{\textsc{GRASP} is fully open source and is available on
GitHub repository at \url{https://github.com/compas/grasp} maintained by the 
\href{https://compas.github.io/}{CompAS collaboration}.}. 
We briefly outline the methods in this section and refer the 
reader to \cite{Grant2007}, \cite{fischer.2016}, \cite{Grasp2018}, and 
\cite{atoms11010007} for details of the MCDHF method and \cite{atoms11040068} for a
step-by-step instruction manual of the package. 

In the MCDHF method, the atomic state wave functions (ASFs, $\Psi(\gamma^{(j)}\, P JM))$ are 
expanded over a linear combination of configuration state functions (CSFs, $\Phi(\gamma_i\, P JM))$:
\begin{equation}
    \Psi(\gamma^{(j)}\, P JM) = \sum_i^{{N_\mathrm{CSFs}}} c^{(j)}_i\, \Phi(\gamma_i\, P JM).
    \label{eq:csfs}
\end{equation}
The CSFs are $jj$-coupled multi-electron functions built from products of one-electron Dirac orbitals.
The radial parts of the one-electron orbitals and the expansion coefficients of the CSFs are 
obtained in a relativistic self-consistent field procedure by solving the Dirac--Hartree--Fock 
radial equations and the configuration interaction eigenvalue problem. 
The angular integrations needed for the construction of the energy functional are based on the 
second quantisation method in the coupled tensorial form \citep{1997JPhB...30.3747G,2001CoPhC.139..263G}. 
Once the radial components of the one-electron orbitals are determined, in the following RCI 
calculations, higher order interactions, such as the Breit interaction and quantum electrodynamic
effects (self-energy and vacuum polarization), are added to the Dirac--Coulomb Hamiltonian.
Keeping the radial components fixed, the expansion coefficients of the CSFs for the target states
are obtained by solving the configuration interaction eigenvalue problem.

Once the ASFs have been determined from RCI calculations, the radiative E1 transition data 
(e.g. transition rates $A$ and weighted oscillator strengths log($gf$)) between two states, 
$\gamma'P'J'$ and $\gamma PJ$, can be computed in terms of reduced matrix elements of the 
transition operator by summing up reduced matrix elements between all the CSFs 
for the lower and upper states:
\begin{eqnarray}
\langle \,\Psi(\gamma PJ)\, \|  {\bf T}^{(1)} \| \,\Psi(\gamma' P'J')\, \rangle =
 \sum_{j,k} c_jc'_k\;\langle \,\Phi(\gamma_j PJ)\, \|  {\bf T}^{(1)} \| \,\Phi(\gamma'_k P'J')\, \rangle.
 \label{eq:tr}
\end{eqnarray}
The transition operator can be expressed in the Babushkin and Coulomb gauges,
which in the non-relativistic limit, respectively, correspond to the length and velocity forms.
In the limit of including CSFs obtained by all possible excitations to a complete set of orbitals, we may arrive at the exact wave functions to the Dirac
equation; this leads to identical values for the Babushkin and Coulomb  transition moments \citep{1974JPhB....7.1458G}.
However, when approximate multi-electron wave functions are used in practice,
the transition data in the two gauges are often different.
The relative difference between the values of these two gauges, d$T$ = ${|A_B-A_C|}/{\mathrm{max}(A_B, A_C)}$, can be used 
as one indicator of the 
accuracy of the wave functions \citep{Fischer2009,Ekman2014}.
It should be noted that d$T$ should be used in a statistical manner 
as done in \cite{2019A&A...621A..16P} and \cite{2023ApJS..265...26L,2023A&A...674A..54L}. 
We discuss this further in Section \ref{sec:accuracy}.

In the MCDHF relativistic calculations, the ASFs are given in terms of $jj$-coupled CSFs. 
In order to identify the computed states and adapt the labelling conventions followed by the
astronomers and experimentalists, the ASFs are transformed from $jj$-coupling to a
basis of $LS\!J$-coupled CSFs. 
In the GRASP2018 package, the transformation from $jj$- to $LS\!J$-coupling is done by the \texttt{jj2lsj} 
program developed by \cite{2003ADNDT..84...99G,2004CoPhC.157..239G,2017Atoms...5....6G}.

\subsection{Computational schemes}
We targeted the states belonging to the \{$\mathrm{3p^3np~(n=3-7)}$, 
${\mathrm{3p^3nf~(n=4,5)}}$\} even configurations and the \{$\mathrm{3s3p^5}$, 
$\mathrm{3p^3ns~(n=4-7)}$, $\mathrm{3p^3nd~(n=3-6)}$\} odd configurations.
To account for higher order configuration--interaction contributions in the wave functions 
relative to the target configurations, we employed extended multi-reference (MR) 
configurations in both MCDHF (MR-MCDHF) and RCI (MR-RCI) calculations.
These extended MR sets were determined through the \texttt{rcsfmr} program in 
\texttt{GRASP2018} based on the results of a preliminary calculation.
The final MR sets adopted in the MCDHF and RCI calculations, the orbital sets, and the total
number of CSFs ($N_\mathrm{CSFs}$) for even and odd parities are presented in Table \ref{tab:MR}. 

Our calculations were performed in the extended optimal level scheme \citep{EOL} for the 
weighted average of the even and odd parity states. 
The multi-reference single-double (MR-SD) method was employed to obtain CSF expansions, 
which allows for single and double (SD) substitutions from MR configurations to orbitals within 
an active set (AS;~\citealt{olsen.1988, sturesson.2007, fischer.2016}).
The orbitals in the AS are divided into spectroscopic orbitals, which build the configurations 
in the MR, and correlation orbitals, which are introduced to correct the initially 
obtained wave functions.
$\mathrm{1s^22s^22p^6}$ was always kept as an inactive closed core.
In the MCDHF calculations, the 3s orbitals were kept inactive and the CSF expansions were 
produced by SD substitutions from all the other valence orbitals of the configurations 
in the MR-MCDHF.
The final wave functions of the target states were determined in a subsequent RCI calculation, 
which included CSF expansions that were formed by allowing SD substitution from the $\mathrm{n\ge3}$ sub-shells
of the MR-MCDHF configurations and single substitution from the $\mathrm{n\ge3}$ sub-shells of the additional
configurations in the MR-RCI to the active sets of orbitals presented in Table \ref{tab:MR}.
Breit interactions were also taken into account in the RCI calculation.
The final even and odd state expansions, respectively, contained 11,049,420 and
8,146,099 CSFs, distributed over different $J$ symmetries
(0--5 for even parity and 0--4 for odd parity).

\subsection{Fine-tuning} \label{sec:finetune}
In atomic-structure calculations, fine-tuning or adjustment of the diagonal elements of the 
Hamiltonian matrix based on experimental energies is a powerful 
semi-empirical technique that can substantially reduce the 
deviations between theoretical and experimentally observed energy levels. 
This technique has been implemented in the \texttt{CIV3}~\citep{HIBBERT1975141} program based on 
the CI method and in the \texttt{ATSP2K}~\citep{FROESEFISCHER2007559} program 
based on the MCHF method. 
It has been successfully applied to the calculations of several atomic systems to obtain 
accurate energy levels and transition parameters 
\citep[e.g.][]{PhysRevA.63.032505,CORREGE200419,FROESEFISCHER20041}.
Both of the above methods are based on $LS\!J$-coupled configuration state functions, 
where the off-diagonal matrix elements between different $LS$-terms are small. 
However, applying fine-tuning directly to $jj$-coupled calculations is challenging due 
to the large off-diagonal matrix elements.
To address this issue, \cite{atoms11040070} proposed a method to convert the Hamiltonian 
from $jj$-coupling to one in $LS\!J$-coupling, where fine-tuning can be applied, and return to the $jj$-coupling form.
The procedure was successfully implemented in \texttt{GRASP2018} through two new 
programs \texttt{jj2lsj\_2022} and \texttt{rfinetune}. 
They also demonstrated that fine-tuning, using C III and B I as examples, 
significantly improved the accuracy of transition rates and lifetimes, 
showing clear improvements compared to \ab calculations.
The $LS$-composition analysis of the ab initio results reveals that, 
with a few exceptions in $\mathrm{3p^3(^4S)7p~^3P}$ and $\mathrm{3s3p^5~^3P}$ states, 
most states in \si can be well described with relatively pure $LS\!J$-coupling. 
Therefore, we adopted the same approach as that employed in \cite{atoms11040070}.

In this work, we applied our fine-tuning procedure for the even and odd states using the MR-RCI set, 
based on experimental energies from the Atomic Spectra Database 
of the National Institute of Standards and Technology (NIST ASD) \citep{NIST_ASD}.
It should be noted that while we were preparing the paper, NIST ASD 
updated the energy levels based on the results from \cite{2024ApJS..274...32C}. Upon careful comparison,
 we found that the corrections on the energy levels were negligible, with most relative differences
  being of the order of 10$^{-7}$. The largest correction occurred in the 3p$^3$($^4$S)7p ~$^5$P$_3$, 
 with a relative difference of 10$^{-4}$. Such small differences will not have significant impacts on
 our fine-tuning results. Therefore, we retained the fine-tuning results in this work.

\section{Results and discussions} \label{sec:results}
\subsection{Energy levels}\label{sec:energy}
\begin{figure}
    \centering
    \includegraphics[width=0.48\textwidth,clip]{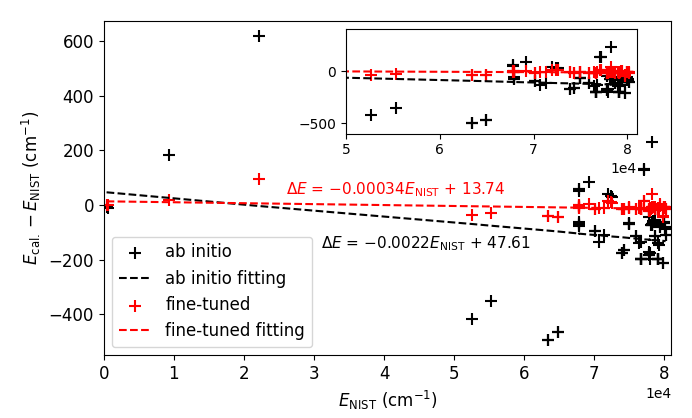}
    \caption{Energy differences as a function of excitation energies provided by NIST ASD. The dashed black and red 
lines are the linear fit to the data from \ab (black plus) and fine-tuning (red plus) calculations, respectively. 
The inset in the upper right shows an enlarged view for $E_\mathrm{NIST} > 5 \times 10^{4}~\mathrm{cm}^{-1}$.}
    \label{fig:energy}
\end{figure}

\begin{figure*}[t]
    \centering
    \includegraphics[width=1.0\textwidth,clip]{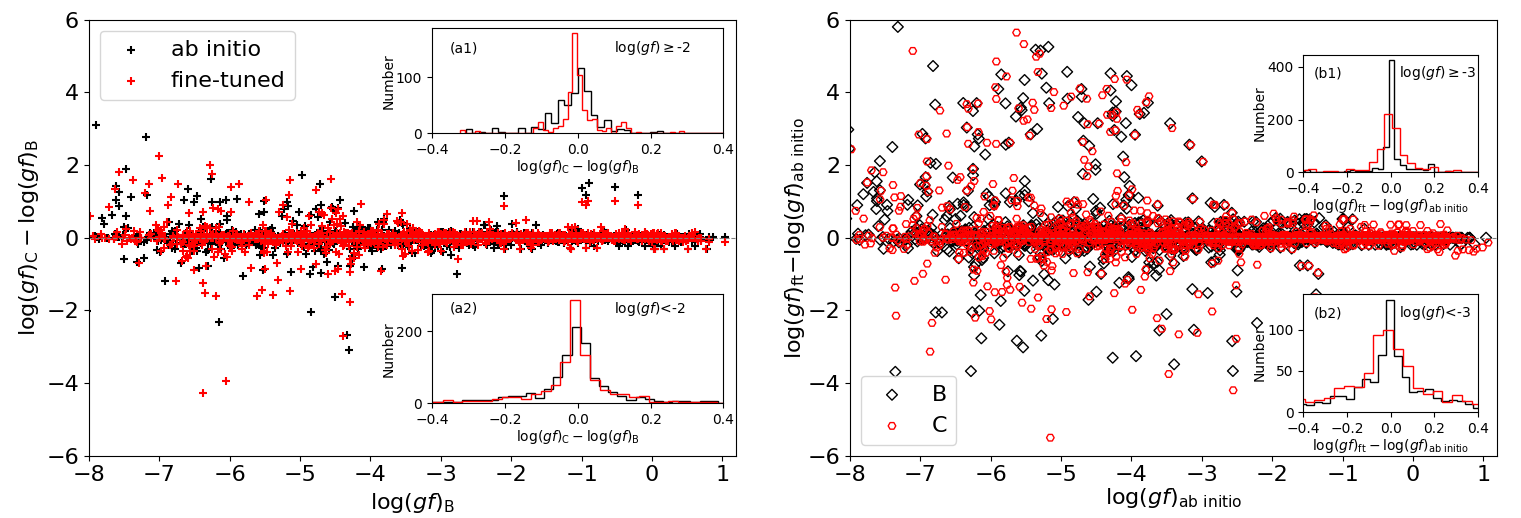}
    \caption{Left: Comparison of log($gf$) between Babushkin and Coulomb gauges for \ab\ (black plus) and \ft\  (red plus) results, respectively.
    Insets (a1) and (a2) show the histograms of the log($gf$) differences for transitions with log($gf$) $\geq -2$ and log($gf$)  < $-2$, respectively.
Right: Comparison of log($gf$) between \ab and \ft calculations for Babushkin (black diamond) and 
Coulomb gauges (red hexagon), respectively.
Insets (b1) and (b2) show the histograms of the log($gf$) differences for transitions with log($gf$) $\geq -3$ and log($gf$)  < $-3$, respectively.}
    \label{fig:gf}
\end{figure*}

The energy levels for the 107 lowest states of \si (56 even states and 51 odd states)
from both \ab and fine-tuning calculations are presented in Table \ref{tab:energy}, 
together with the experimental energies from the recent update of the NIST ASD by \cite{2024ApJS..274...32C}.
The numbers in the first column of the table, labelled as `No.' are in the order of the energy levels given in the NIST ASD.
Figure \ref{fig:energy} shows the energy differences between calculated data and NIST ASD values 
plotted against the excitation energies, $E_\mathrm{NIST}$. 
The agreement between the computational and experimental energies is systematically improved after fine-tuning; 
the systematic error of computational excitation energies decreases from 0.22\% to 0.03\%, and the root-mean-square 
deviation decreases from 188.5 cm$^{-1}$ to 20.3 cm$^{-1}$.
Especially for the energy levels of $\mathrm{3p^4~^1S_0}$, $\mathrm{3p^3(^4S)4s~^5S^o_2, ^3S^o_1}$, and 
$\mathrm{3p^3(^4S)4p~^5P_{1,2,3}, ^3P_{0,1,2}}$, the \ab energies are rather far away (more than 
300 cm$^{-1}$) from the experimental energies; in contrast, the \ft energies are all within 50 cm$^{-1}$ of the experimental values, 
except for $\mathrm{3p^4~^1S_0}$, which differs by 96 cm$^{-1}$.
In addition, fine-tuning can help correct the energy positions of the levels, where the positions refer to
the level locations arranged in terms of the excitation energies.
For example, in the \ab calculation, the $\mathrm{3p^3(^2P)4s~^1P^o_1}$ state is higher than 
the $\mathrm{3p^3(^2D)4p~^3F}$ states 
and in the wrong position compared to experimental data.
Fine-tuning brings the energy into the correct position with a corresponding 
slightly decreased mixing with the $\mathrm{3p^3(^2D)3d~^1P^o_1}$ perturber 
into the $\mathrm{3p^3(^2P)4s~^1P^o_1}$ state.

Further iterations of fine-tuning could potentially yield results even closer to the experimental energies. 
Considering the total time cost involved in the fine-tuning process and the recalculation of the wave functions and transition properties,
as well as the fact that currently available results have already 
reached a level of accuracy that is challenging to attain simply by increasing the 
size of the CSF basis, especially given that the present work involved over ten million CSFs in the calculations, 
we believe that the current results are sufficient to study 
the impact of fine-tuning on transition data.

\subsection{Transition data}\label{sec:tr}

The transition parameters computed from both \ab and fine-tuning calculations for 1,730 E1 
transitions among the 107 energy levels given in Table~\ref{tab:energy} are presented in 
Table~\ref{tab:tr}; the full table is available at the CDS.
Note that the wavelengths listed in all tables of the paper are derived 
from the experimental energy levels in the NIST ASD, which are originally from \cite{2024ApJS..274...32C}; the transition data, including
weighted oscillator strengths, log($gf$) 
and transition rates, $A$, for both \ab and \ft results are all adjusted 
using experimental wavelengths.

\begin{figure*}[t]
    \centering
    \includegraphics[width=1.0\textwidth,clip]{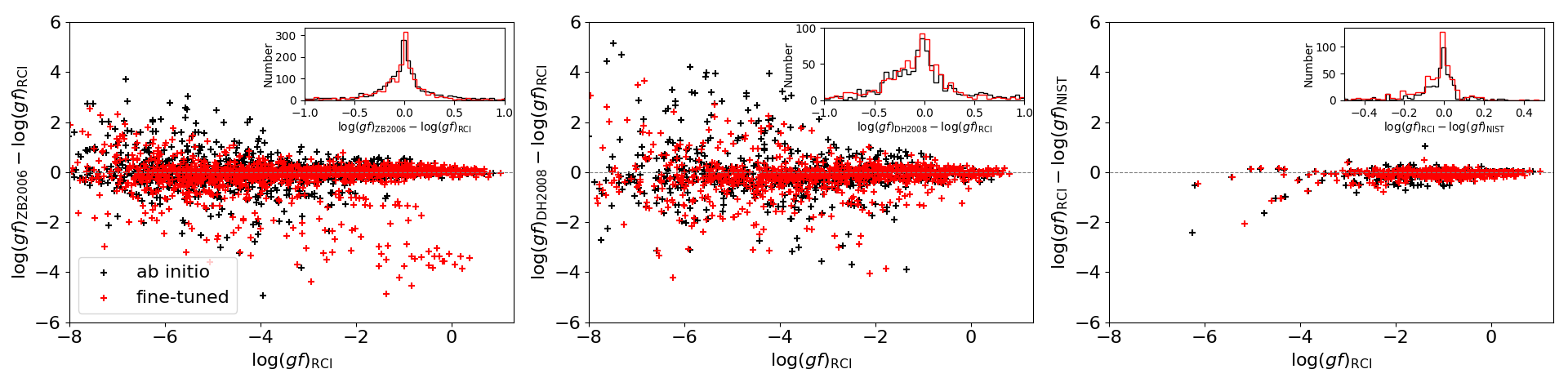}
    \caption{Comparison of log($gf$) from the current study with those from previous works. 
    Left: ZB2006 \citep{2006JPhB...39.2861Z}.
    Middle: DH2008 \citep{2008ADNDT..94..561D}.
    Right: NIST ASD \citep{NIST_ASD}. The data in NIST ASD were compiled 
    from various sources, including \cite{1968ZNatA..23.1707M,1994ApJ...428..393B,1996A&A...309..991B,1998ApJ...502.1010B,RZerne_1997,FROESEFISCHER2006607} 
    and \cite{2006JPhB...39.2861Z}. 
    Babushkin results from present \ab (black plus) and \ft (red plus) RCI calculations are
    used in the comparison. The inset figures show the histograms of the distribution of the number of transitions for the differences in log($gf$). 
    Note that the y-axis range is consistent across all plots in Figures
    \ref{fig:gf} and \ref{fig:tr}.}
    \label{fig:tr}
\end{figure*}

\subsubsection{Accuracy assessment}\label{sec:accuracy}

We evaluated the accuracy of our computed transition data following the method
used in \cite{2023A&A...674A..54L}.  This approach is based on grouping
transitions based on the magnitude of $A$ values.  For \si, we chose six groups
such that $A$ < 10$^{-2}$ s$^{-1}$, 10$^{-2}$ $\le A$ < 10$^{0}$ s$^{-1}$,
10$^{0}$ $\le A$ < 10$^{3}$ s$^{-1}$, 10$^{3}$ $\le A$ < 3.5$\times$10$^{5}$
s$^{-1}$, 3.5$\times$10$^{5}$ $\le A$ < 10$^{7}$ s$^{-1}$, and $A \ge$ 10$^{7}$.
Next, we calculated the averaged d$T_\mathrm{av}$ for each group.  As we
mentioned in Section \ref{sec:MCDHF}, the gauge difference parameter should only
be used as an uncertainty estimate in a statistical manner for a group of
transitions with similar properties. As such, we define d$\tilde{T}$ as
max(d$T$, d$T_\mathrm{av}$) to replace d$T$ as the gauge difference indicator
for individual transitions.  We then folded the cancellation-factor (CF) into the analysis.
This is a numerical measure of the cancellation effect in the
computation of transition matrix elements
\citep[e.g.][]{Cowan_book,2013A&A...551A,2020ApJS..248...13G}.  A small CF
(generally under about 0.1 or 0.05 as given in \cite{Cowan_book})
indicates a strong cancellation effect, resulting in computed transition
parameters with large uncertainties.

The final accuracy class for each computed transition rate was estimated by a
combined analysis of the d$\tilde{T}$ and CF parameters.  Based on the above
procedure, we divided the accuracy into five classes, following the definition conventions 
of the accuracy classification in the NIST ASD. The A, B, C, D, and E classes correspond to 
the \{A, A+, and AA\}, \{B+ and B\}, \{C+ and C\}, \{D+ and D\}, 
 and E classes, respectively, as defined by the NIST ASD. The definitions of
d$\tilde{T}$ and CF for each accuracy class are as follows:
\begin{itemize}
\item A ($\le$ 3\%): d$\tilde{T}$ $\le$ 3 \& CF $\ge$ 0.1
\item B ($\le$ 10\%): 3 $<$ d$\tilde{T}$ $\le$ 10 \& CF $\ge$ 0.1 or \\
\hspace*{70pt} d$\tilde{T}$ $\le$ 3 \& CF $<$ 0.1
\item C ($\le$ 25\%): 10 $<$ d$\tilde{T}$ $\le$ 25 \& CF $\ge$ 0.1 or \\
\hspace*{54pt} 3 $<$ d$\tilde{T}$ $\le$ 10 \& CF $<$ 0.1
\item D ($\le$ 50\%): 25 $<$ d$\tilde{T}$ $\le$ 50 \& CF $\ge$ 0.1 or \\
\hspace*{49pt} 10 $<$ d$\tilde{T}$ $\le$ 25 \& CF $<$ 0.1
\item E ($>$ 50\%): d$\tilde{T}$ $>$ 50 or \\
\hspace*{48pt} 25 $<$ d$\tilde{T}$ $\le$ 50 \& CF $<$ 0.1
\end{itemize}

We provide the estimated accuracy class for each transition in Table \ref{tab:tr}; 
these were defined based on the above procedure for both \ab and \ft results.
We hope that this information can provide data users with a reference for data quality.
Taking the Babushkin gauge as an example, the percentage fraction of each 
accuracy class is as follows:
\begin{itemize}
\item \ab: A: 2.1\%; B: 13.8\%; C: 30.0\%; D: 28.6\%; E: 25.5\%
\item \ft: A: 9.1\%; B: 15.1\%; C: 26.1\%; D: 26.8\%; E: 22.9\%
\end{itemize}
The reason for the high proportion of lower accuracy data is that among the 1,730
transition data, approximately 50\% are $LS$-forbidden inter-combination 
transitions (IC), including two-electron one-photon (TEOP) transitions involving
the $\mathrm{3s3p^5}$ energy levels. These transitions are governed by electron 
correlation effects, which are difficult to calculate accurately. 
In high-accuracy classes A and B, over 80\% of the transitions are 
$LS$-allowed, whereas in the low-accuracy classes D and E, more than 70\% are 
$LS$-forbidden IC transitions.

The statistical results indicate that the proportion of high-accuracy results 
in classes A and B obtained from the fine-tuned calculations is higher than that
from the \ab calculations. The primary reason for this difference is that 
fine-tuning reduces the overall differences in the transition data
between the Babushkin and Coulomb gauges.

\subsubsection{Comparison between Babushkin and Coulomb gauges}\label{sec:BC}
The left panel of Figure \ref{fig:gf} presents the differences in log($gf$) values 
between the Babushkin and Coulomb gauges. 
The histograms (insets (a1) and (a2)) show the distribution of the gauge difference for 
transitions with log($gf$) $\geq -2$ and log($gf$) < $-2$, respectively.
Both the \ab and \ft data show good agreement between the two gauges, 
with 73\% of the total of 1,730 transitions differing by less than 0.1 dex. 
For transitions with log($gf$) greater than $-$2.0, 
approximately 86\% agree within 0.1 dex between the two gauges for both \ab and \ft results.
In the same subset, as can be seen from the histogram (a1) in the left panel of Figure \ref{fig:gf}, 
the number of transitions with differences between the two gauges greater than 0.05 dex is higher for the \ab
results (33\%) than for the \ft results (23\%).
This indicates that fine-tuning can enhance the consistency between the Babushkin 
and Coulomb gauges to a certain degree, which accounts for the higher proportion of high-accuracy
 data in the fine-tuned results compared to those obtained from ab initio 
calculations, as demonstrated in Section \ref{sec:accuracy}.

Generally, stronger transitions show better agreement between the Babushkin and Coulomb gauges. 
However, we observed six outliers in the range of $-$1 to 0 of log($gf$)$\mathrm{_B}$ in Figure \ref{fig:gf}.
These are transitions between high Rydberg states 
$\mathrm{3p^3(^4S)7p~^3P_{0,1,2}}$ and $\mathrm{3p^3(^4S)6d~^3D^o_{1,2,3}}$. 
Despite the large differences (by 1--1.5 dex) between the Babushkin and Coulomb 
gauges for these transitions, 
the \ab and \ft data in the Babushkin gauge show excellent agreement, 
with deviations of less than 0.01 dex, 
while the results in the Coulomb gauge differ by 0.2--0.5 dex.
Although all six transitions are classified as E
accuracy class based on the method described in Section \ref{sec:accuracy}, we
suggest that the Babushkin gauge data may be more reliable for these
transitions.

\subsubsection{Comparison between \ab and \ft results}\label{ab-ft}

The right panel of Figure \ref{fig:gf} presents a comparison of log($gf$) values 
between the fine-tuned and the ab initio results. The histograms (insets (b1) and (b2)) display the distribution 
of the differences for transitions with log($gf$) $\geq -$3 and log($gf$) $< -$3, respectively.
It is important to note that all data, including wavelengths and transition data, 
were adjusted to match the experimental energy levels, ensuring that the differences shown 
in the figure are solely due to the changes in mixing coefficients caused by fine-tuning. 
From this comparison, we observe that fine-tuning has a significant impact on 
log($gf$) values, with approximately 42\% of the Babushkin data 
(and 45\% of the Coulomb gauge data) showing changes greater than 0.1 dex.
Second, as can be seen from the histogram (b1), the effect of fine-tuning is more pronounced on the Coulomb gauge results than on the 
Babushkin gauge data.
For example, among transitions with log($gf$) greater than $-$3.0, 
42\% of the Coulomb data show changes greater than 0.05 dex,
compared to 26\% of the Babushkin data.
Third, fine-tuning has a more substantial effect on weak transitions 
than on strong transitions.
The histograms show that 44\% of transitions with $\log(gf) < -3$  have a change in the Babushkin 
gauge greater than 0.2 dex, compared to 9\% of transitions with $\log(gf) \geq -3$.

Despite ongoing debates concerning the choice of the appropriate gauge \citep[e.g.][]{PhysRevA.3.1242, Hibbert_1974,Cowan_book,Papoulia2019,2024MNRAS.530.5220G}, 
the Babushkin gauge is generally considered to yield more reliable transition results, as it better describes the 
outer part of the wave functions that governs the atomic transitions \citep{Hibbert_1974}. 
In addition, the contributions from the negative-energy states were not taken into 
account in the MCDHF and RCI calculations, which have
negligible effects on the Babushkin gauge results but can affect the Coulomb 
gauge results significantly, especially for the weak 
IC transitions; this further suggests that the Babushkin gauge results
appear to be more reliable \citep{2001PhRvA..64d2507C}.
This explains the findings presented in Figure \ref{fig:gf} and the conclusions in Section \ref{sec:accuracy} and \ref{sec:BC}, 
which namely show that fine-tuning can better improve the less accurate results in the Coulomb gauge,
thereby enhancing the agreement between the Babushkin and Coulomb gauges and 
consequently improving the overall accuracy.
Furthermore, the transition matrix element in Coulomb gauge contains a dependence on the transition energy, making
it more sensitive to fine-tuning.
For these reasons, the greater impact of fine-tuning on the Coulomb gauge 
than on the Babushkin gauge is expected.

\subsubsection{Comparison with previous theoretical and experimental results}
\label{transition_comparison_previous}

\begin{figure}
    \centering
    \includegraphics[width=0.45\textwidth,clip]{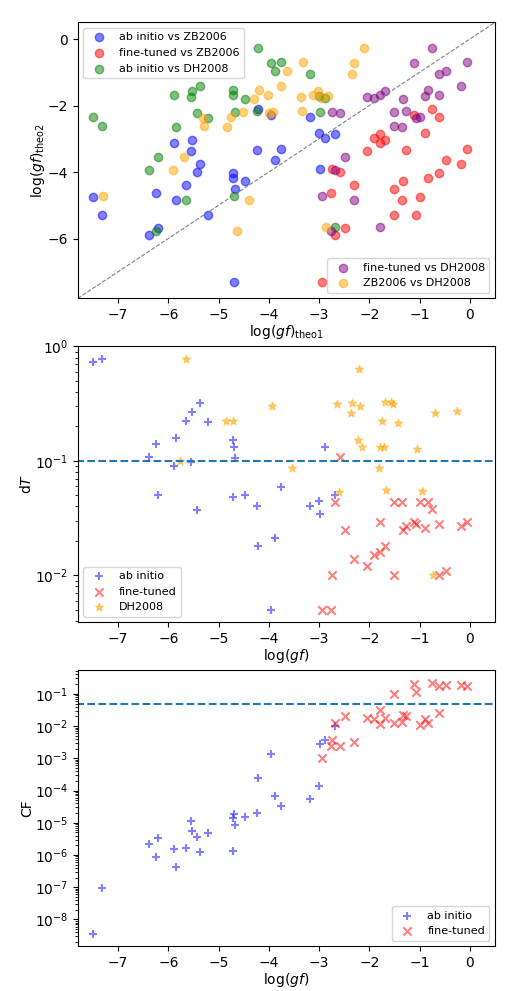}
    \caption{Upper panel: Comparison of theoretical log($gf$) values for transitions with log($gf$)$\mathrm{_{RCI}}$ > $-$3.0 
    and $\Delta$log($gf$) > 1.0 dex shown in the left and middle panels of Figure \ref{fig:tr}. 
    Middle panel: Scatter plot of d$T$ values. Lower panel: Scatter plot of CF values. 
   The dashed lines in the middle panel and the lower panel represent d$T$ = 0.1 and CF = 0.05, respectively.
    ZB2006: \cite{2006JPhB...39.2861Z}; DH2008: \cite{2008ADNDT..94..561D}. }
    \label{fig:diff}
\end{figure}

\begin{table*}
\small
\caption{\label{tab:loggf} Comparison of experimental and theoretical weighted oscillator strength, log($gf$), for \si{}.}
\centering
\begin{tabular}{llccccccccccccccc}
\hline\midrule
     \multirow{2}{*}{Upper State}    &   \multirow{2}{*}{Lower State}   &   $\lambda_{vac.}$ & \multicolumn{2}{c}{ab initio} & \multicolumn{2}{c}{Fine-tuned}   &  \multirow{2}{*}{ZB2006$^a$}  & \multicolumn{2}{c}{DH2008$^b$} &  \multirow{2}{*}{Experiments}  \\
\cmidrule{4-5}\cmidrule{6-7}\cmidrule{9-10}
                                     &                                  &    (\AA) &  B & C &      B      &    C   &   &  B     &   C    &       \\ 
\midrule
3p$^3$($^2$P)4s ~$\mathrm{^3P^o_2}$ &  3p$^4$($^3$P)       ~$\mathrm{^3P_2}$    &  1295.653  &  $-$0.389  &   $-$0.397   &   $-$0.391   & $-$0.398 &  $-$0.385   &   $-$0.412  &    $-$0.434 &  $-$0.362$\pm$0.030$^c$  \\
3p$^3$($^2$P)4s ~$\mathrm{^3P^o_1}$ &  3p$^4$($^3$P)       ~$\mathrm{^3P_2}$    &  1296.174  &  $-$0.874  &   $-$0.882   &   $-$0.878   & $-$0.886 &  $-$0.870   &   $-$0.888  &    $-$0.907 &  $-$0.959$\pm$0.039$^c$  \\
3p$^3$($^2$P)4s ~$\mathrm{^3P^o_2}$ &  3p$^4$($^3$P)       ~$\mathrm{^3P_1}$    &  1302.336  &  $-$0.801  &   $-$0.809   &   $-$0.801   & $-$0.808 &  $-$0.805   &   $-$0.827  &    $-$0.845 &  $-$0.815$\pm$0.043$^c$  \\
3p$^3$($^2$P)4s ~$\mathrm{^3P^o_1}$ &  3p$^4$($^3$P)       ~$\mathrm{^3P_1}$    &  1302.862  &  $-$1.078  &   $-$1.087   &   $-$1.078   & $-$1.086 &  $-$1.074   &   $-$1.084  &    $-$1.103 &  $-$0.932$\pm$0.067$^c$  \\
3p$^3$($^2$P)4s ~$\mathrm{^3P^o_1}$ &  3p$^4$($^3$P)       ~$\mathrm{^3P_0}$    &  1305.883  &  $-$0.910  &   $-$0.919   &   $-$0.910   & $-$0.917 &  $-$0.914   &   $-$0.921  &    $-$0.939 &  $-$0.824$\pm$0.038$^c$  \\
3p$^3$($^4$S)4s ~$\mathrm{^3S^o_1}$ &  3p$^4$($^3$P)       ~$\mathrm{^3P_2}$    &  1807.312  &  $-$0.367  &   $-$0.374   &   $-$0.370   & $-$0.379 &  $-$0.371   &   $-$0.369  &    $-$0.394 &  $-$0.319$\pm$0.032$^c$  \\
3p$^3$($^4$S)4s ~$\mathrm{^3S^o_1}$ &  3p$^4$($^3$P)       ~$\mathrm{^3P_1}$    &  1820.342  &  $-$0.597  &   $-$0.605   &   $-$0.600   & $-$0.610 &  $-$0.601   &   $-$0.599  &    $-$0.624 &  $-$0.593$\pm$0.056$^c$  \\
3p$^3$($^4$S)4s ~$\mathrm{^3S^o_1}$ &  3p$^4$($^3$P)       ~$\mathrm{^3P_0}$    &  1826.245  &  $-$1.077  &   $-$1.084   &   $-$1.080   & $-$1.089 &  $-$1.080   &   $-$1.078  &    $-$1.103 &  $-$1.252$\pm$0.085$^c$  \\
\midrule
\multicolumn{10}{c}{Abundance diagnostic \si lines} \\
\midrule
3p$^3$($^4$S)5p ~$\mathrm{^5P_3}$   &  3p$^3$($^4$S)4s    ~$\mathrm{^5S^o_2}$  &  4695.422  &  $-$1.667  &   $-$1.690   &   $-$1.675   & $-$1.686 &  $-$1.716   &   $-$1.599  &    $-$1.607 & \\ 
3p$^3$($^4$S)5p ~$\mathrm{^5P_2}$   &  3p$^3$($^4$S)4s    ~$\mathrm{^5S^o_2}$  &  4696.753  &  $-$1.825  &   $-$1.849   &   $-$1.833   & $-$1.845 &  $-$1.873   &   $-$1.753  &    $-$1.757 & \\ 
3p$^3$($^4$S)5d ~$\mathrm{^5D^o_2}$ &  3p$^3$($^4$S)4p    ~$\mathrm{^5P_3}$    &  6758.611  &  $-$1.858  &   $-$1.896   &   $-$1.857   & $-$1.869 &    $-$1.784       &           &           & \\ 
3p$^3$($^4$S)5d ~$\mathrm{^5D^o_3}$ &  3p$^3$($^4$S)4p    ~$\mathrm{^5P_3}$    &  6758.886  &  $-$1.014  &   $-$1.051   &   $-$1.012   & $-$1.024 &    $-$0.940       &           &           & \\ 
3p$^3$($^4$S)5d ~$\mathrm{^5D^o_4}$ &  3p$^3$($^4$S)4p    ~$\mathrm{^5P_3}$    &  6759.043  &  $-$0.428  &   $-$0.465   &   $-$0.427   & $-$0.438 &  $-$0.353   &           &           & \\ 
3p$^3$($^4$S)4d ~$\mathrm{^5D^o_0}$ &  3p$^3$($^4$S)4p    ~$\mathrm{^5P_1}$    &  8672.585  &  $-$0.911  &   $-$0.962   &   $-$0.918   & $-$0.932 &  $-$0.908   &   $-$0.826  &    $-$0.873 & \\ 
3p$^3$($^4$S)4d ~$\mathrm{^5D^o_4}$ &  3p$^3$($^4$S)4p    ~$\mathrm{^5P_3}$    &  8697.023  &   +0.047  &   $-$0.004   &    +0.040   &  +0.027 &   +0.052   &    +0.186  &     +0.182 & \\ 
3p$^3$($^4$S)4p ~$\mathrm{^3P_2}$   &  3p$^3$($^4$S)4s    ~$\mathrm{^3S^o_1}$  &  10458.308 &   0.259  &    +0.273   &    +0.259   &  +0.262 &   +0.256   &    +0.270  &     +0.274 &   +0.250$\pm$0.009$^d$  \\
3p$^3$($^4$S)4p ~$\mathrm{^3P_0}$   &  3p$^3$($^4$S)4s    ~$\mathrm{^3S^o_1}$  &  10459.618 &  $-$0.440  &   $-$0.426   &   $-$0.440   & $-$0.436 &  $-$0.444   &   $-$0.429  &    $-$0.426 &  $-$0.447$\pm$0.011$^d$  \\
3p$^3$($^4$S)4p ~$\mathrm{^3P_1}$   &  3p$^3$($^4$S)4s    ~$\mathrm{^3S^o_1}$  &  10462.267 &   +0.037  &    +0.051   &    +0.037   &  +0.041 &   +0.035   &    +0.048  &     +0.052 &   +0.030$\pm$0.010$^d$  \\
\bottomrule
\end{tabular}
\tablefoot{
B: Babushkin gauges. C: Coulomb gauges. The vacuum wavelengths are used in the table.\\
\tablefoottext{a}{\cite{2006JPhB...39.2861Z}.}
\tablefoottext{b}{\cite{2008ADNDT..94..561D}.}
\tablefoottext{c}{Beam-foil technique by \cite{1994ApJ...428..393B}.}
\tablefoottext{d}{Laser spectroscopy by \cite{RZerne_1997}.}
}
\end{table*}

In Figure \ref{fig:tr}, our ab initio and fine-tuned log($gf$) results are compared with 
the data from ZB2006 (1719 transitions), DH2008 (1001 transitions), and 
NIST ASD\footnote{Note that some log($gf$) values for \si{} transitions in 
NIST ASD are anomalous (Alexander Kramida, private communication), while the 
line strength, $S$, and oscillator strength, $f$, values are correct. 
The log($gf$) values from NIST ASD adopted in this paper were all calculated from the $f$ values.} (528 transitions). 
ZB2006 is based on the B-spline box-based multi-channel method, 
while DH2008 is based on CI methods with fine-tuning. 
The histograms in each panel show the statistical distribution of the differences between
our results and the others.
Overall, our results show better agreement with ZB2006 than with DH2008. 
For example, among transitions with log($gf$) greater than $-$3.0, 68.4\% (66.9\%) of the 
ab initio (fine-tuned) data agree within $\pm$0.1 dex with ZB2006, 
while only 44.6\% agree with DH2008. 

However, some unexpected large deviations are observed in the lower right corner
of the left panel in Figure \ref{fig:tr}, indicating that fine-tuning leads to large 
discrepancies of $>$1.0 dex compared with ZB2006 data for certain transitions. 
Most of these transitions are IC transitions between $\mathrm{3p^{3}(^{2}D)4p~^{1}P}$, 
$\mathrm{3p^{3}(^{4}S)nd~(n=3,4,5,6)~^{3}D^o,~^{5}D^o}$ and 
$\mathrm{3p^{3}(^{4}S)nf~(n=3,4,5)~^{3}F,~^{5}F}$.
IC transitions are directly caused by the mixing of states, and their theoretical 
transition rates are very sensitive to the quality of the wave functions; 
the accurate calculation of their transition data is more challenging than 
for $LS$-allowed transitions.
In the upper panel of Figure \ref{fig:diff}, we compare the results of the anomalous transitions 
for which both ZB2006 and DH2008 data are available.
As can be seen from the figure, our fine-tuned data are the largest, 
while our ab initio data are the smallest among all the calculated results.
It is also interesting to note that, despite the considerable differences 
among various results,
our fine-tuned data show better agreement with 
the fine-tuned results of DH2008, while the ab initio results are closer to ZB2006. 
Upon closer inspection of the energy levels associated with these `anomalous' transitions, 
we observe no tendency for the fine-tuning process to `over-correct' for inaccuracies in energy separations.
In the middle and lower panels of Figure \ref{fig:tr}, we show the distribution of the d$T$ and CF parameters.
It is clear that the fine-tuned results have better consistency between the Babushkin and Coulomb gauges than the
ab initio and the DH2008 results.
The fine-tuned results exhibit larger CF values than the ab initio data.
Based on the accuracy assessment using the d$\tilde{T}$ \&CF method in Section \ref{sec:accuracy}, 
all transitions from the fine-tuning calculations in Figure \ref{fig:diff} have an accuracy class better than C (i.e. A, B or C), 
while all results from the ab initio calculations are worse than class C (i.e. C, D, or E). 
Therefore, with regard to a d$T$ and CF assessment, 
the fine-tuned results are considered more accurate. 
We hope for precise experimental measurements in the future to verify them.

Of the 1730 transitions presented in this work, data for 528 of them are available in NIST ASD;
they were compiled by \cite{1993JPCRD..22..279K} and \cite{2009JPCRD..38..171P}. 
These data mainly consist of the stronger transitions ($S \geq$ 0.01) from ZB2006, while the remaining transitions were
 sourced from other theoretical calculations \citep{1996A&A...309..991B,1998ApJ...502.1010B,FROESEFISCHER2006607, 2006JPhB...39.2861Z} 
 and experimental measurements \citep{1968ZNatA..23.1707M,1994ApJ...428..393B, RZerne_1997} .
NIST ASD critically compiled and assessed the uncertainties associated with these results. 
A comparison of our results with these compiled data helped us assess the reliability of our calculated results.
A comparison between our calculated results and the data from NIST ASD is presented in the right panel of Figure \ref{fig:tr}.
We can see that our calculated data, 
both ab initio and fine-tuned, agree well with the recommended log($gf$) values from
NIST ASD, with approximately 74\% of the results falling within $\pm$0.1 dex and 
54\% falling within $\pm$0.05 dex.

\begin{figure}
    \centering
    \includegraphics[width=0.49\textwidth,clip]{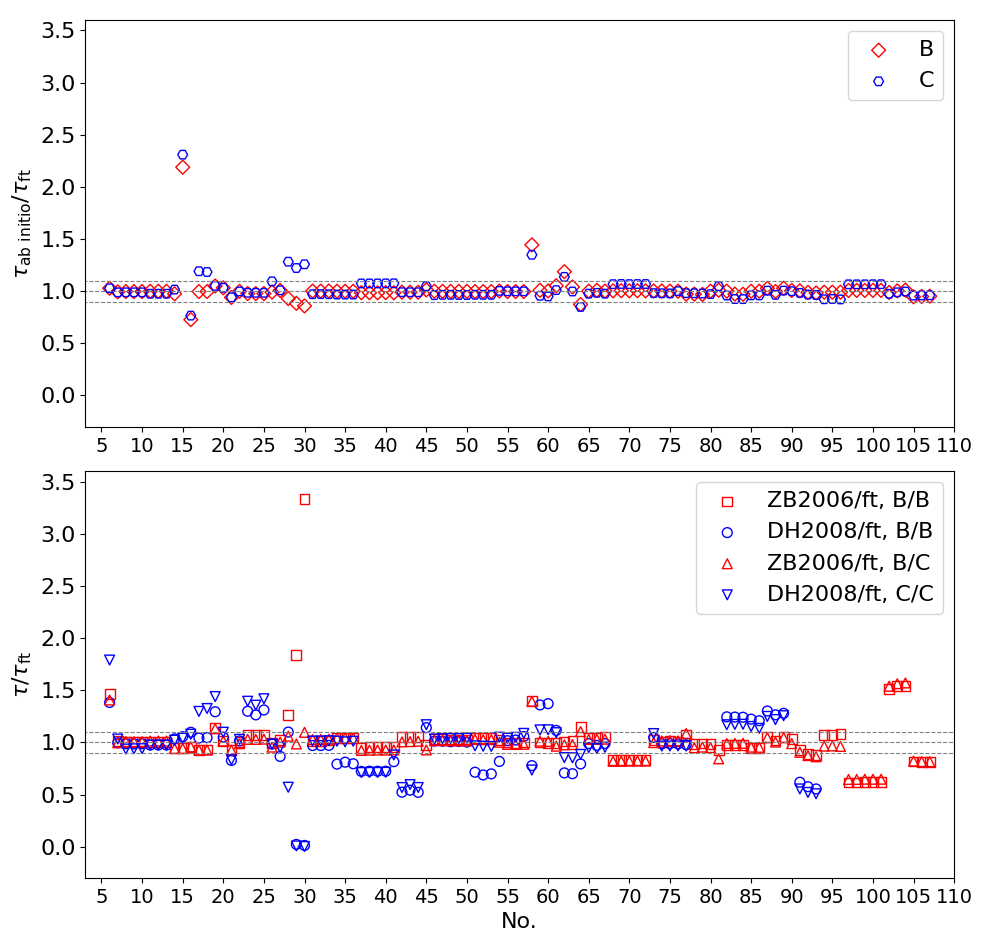}
    \caption{Upper: Comparison of \ft lifetimes (ft) between Babushkin (B) and Coulomb 
    (C) gauges. Lower: Comparison of \ft lifetimes with other theoretical results.
    ZB2006: \cite{2006JPhB...39.2861Z}; DH2008 \cite{2008ADNDT..94..561D}. ZB2006/ft,
    B/B: the Babushkin gauge of ZB2006 divided by the Babushkin gauge of fine-tuned lifetimes.
    The three dashed grey lines represent $\tau/\tau_\mathrm{ft}$ = 0.9, 1.0, and 1.1, respectively.}
    \label{fig:tau}
\end{figure}

Specifically, the results for the transitions with available experimental measurements and 
for the ten transitions used as solar abundance diagnostics are presented 
in Table \ref{tab:loggf}. 
The theoretical results include ab initio and fine-tuned RCI calculations, 
B-splines calculation from ZB2006 and CI calculation from DH2008.
The laboratory measured results from \cite{1994ApJ...428..393B} and  \cite{RZerne_1997}  are listed in the last column.
\cite{1994ApJ...428..393B} derived the oscillator strengths 
based on measured mean lifetimes and branching ratios using beam-foil spectroscopic techniques. 
However, all theoretical results for spectral lines, including 1296.17 \AA, 
1203.86 \AA, 1305.88 \AA, 1807.31 \AA, and 1826.24 \AA, 
fall outside the experimental uncertainties by \cite{1994ApJ...428..393B}, 
while different theoretical calculations agree very well with each other to $\pm$0.03 dex. 
We further examined the measured results reported by \cite{1968ZNatA..23.1707M}, 
which provided the oscillator strengths for individual lines of the 1814 \AA~and 1299 \AA~multiplets.  
As shown in \cite{1994ApJ...428..393B}, the discrepancies between the results of the two experiments are particularly large, 
with a minimum difference of 0.1 dex in log($gf$). This indicates that there may still be considerable 
uncertainties in the measured results.

The comparison results for the ten visible and near-infrared spectral lines
used by \cite{2015A&A...573A..25S} for the abundance analysis of the Sun are shown in the lower half of Table \ref{tab:loggf}. 
For the 10450 \AA{} triplet lines, different theoretical results and laboratory measurements 
based on laser spectroscopy by \cite{RZerne_1997} agree very well within 0.02 dex, 
although the results from DH2008 are slightly outside the measurement uncertainties. 
For all ten transitions, the results from DH2008 are generally larger than those from 
ZB2006 and present calculations, 
while our ab initio and fine-tuned results are in overall good agreement with ZB2006. 
However, for the two 4690 \AA{} lines and the 6750 \AA{} triplet, the differences between our results 
and those of ZB2006 are large, with the maximum difference reaching 0.07 dex. 
These lines were used as diagnostics for the solar sulphur abundance \citep[e.g.][]{2007A&A...470..699C,2015A&A...573A..25S}. 
However, owing to this significant discrepancy, 6750 \AA{} was given zero weight in the recent analysis of \citet{2025A&A...703A..35A}, and requires
for further improvements to the theoretical calculations. Precise laboratory measurements of the 
oscillator strengths for these diagnostic lines are highly desirable for the future, 
not only to better constrain theoretical calculations, but also for astrophysical applications.
Astrophysical consistency tests may also help to assess the reliability and quality of theoretical data \citep[e.g.][]{2019A&A...624A..60L,2024ARep...68.1159K},
especially when laboratory measurements are unavailable.
\begin{table*}
\tiny
\caption{\label{tab:tau} Comparison of experimental and theoretical radiative lifetimes (in units of nanosecond) for \si{}.}
\centering
\begin{tabular}{cccccccccccccccccccc}
\hline\midrule
 & &  \multicolumn{7}{c}{Theory}  && \multicolumn{2}{c}{Experiment} \\
\cmidrule{3-8}\cmidrule{10-11}
  &     &   ab initio  & Fine-tuned       & B-spline$^a$  &  \multirow{2}{*}{HFR$^b$} &     CI$^c$ &   \multirow{2}{*}{MCHF$^d$} && laboratory  & astronomical  \\
No.  & State    & B/C & B/C  & B/C  &  & B/C &  &&  measurements & observations$^f$  \\
\midrule
  7 & 3p$^3$($^4$S)4s ~$\mathrm{^3S_1}$ &  1.93/1.96           &  1.94/1.99       &  1.93/1.98                &  1.5     &  1.94/2.05       &     2.04    &&    1.875(0.094)$^g$&    1.875(0.188) \\
 27 & 3p$^3$($^4$S)5s ~$\mathrm{^3S_1}$ &  6.88/7.01           &  6.83/6.95       &  6.87/6.97                &  4.2     &  5.92/6.63       &             &&      7.1(0.5)$^e$  &    6.9(1.4)     \\
 54 & 3p$^3$($^4$S)6s ~$\mathrm{^3S_1}$ &  16.80/17.20         &  16.77/17.14     &  17.0/17.2                &  9.7     &  13.72/18.00     &             &&      17.7(1.1)$^e$ &    16.9(3.4)    \\
 90 & 3p$^3$($^4$S)7s ~$\mathrm{^3S_1}$ &  34.16/35.38         &  33.90/35.38     &  34.5/34.9                &  18.8    &                  &             &&      35.6(4.0)$^e$ &    31.0(8.4)    \\
  6 & 3p$^3$($^4$S)4s ~$\mathrm{^5S_2}$ &  13893/14439         &  13510/14043     &  17615/19771              &          &  18691/25138     &    39037    &&      9200(1000)$^i$        &                 \\
            &                       &                      &                  &                           &          &                  &             &&       27000(5000)$^j$           &                 \\
 22 & 3p$^3$($^2$D)4s ~$\mathrm{^1D_2}$ &  2.23/2.27           &  2.23/2.27       &  2.23/2.24                &  1.7     &  2.26/2.34       &     2.30    &&                &                 \\
 55 & 3p$^3$($^2$P)4s~$\mathrm{^3P_0}$ &   2.25/2.29           &  2.25/2.29       &  2.22/2.25                &  3.0     &  2.29/2.39       &             &&                &                 \\   
 56 & 3p$^3$($^2$P)4s~$\mathrm{^3P_1}$ &   2.24/2.28           &  2.24/2.28       &  2.21/2.24                &  2.9     &  2.30/2.40       &             &&  2.034(0.102)$^g$ &    2.1(0.2)     \\ 
 57 & 3p$^3$($^2$P)4s~$\mathrm{^3P_2}$ &   2.22/2.26           &  2.22/2.26       &  2.20/2.22                &  2.8     &  2.35/2.46       &             &&  2.146(0.129)$^g$ &    2.1(0.2)     \\
 12 & 3p$^3$($^4$S)4p ~$\mathrm{^3P_0}$ &  45.12/43.72         &  45.15/44.80     &  46.3/45.4                &  36.1    &  44.05/43.67     &     49.5    &&      46.1(1.0)$^e$ &                 \\
 11 & 3p$^3$($^4$S)4p ~$\mathrm{^3P_1}$ &  45.14/43.74         &  45.17/44.82     &  46.3/45.4                &  36.1    &  44.04/43.66     &     49.4    &&      46.1(1.0)$^e$ &                 \\
 13 & 3p$^3$($^4$S)4p ~$\mathrm{^3P_2}$ &  45.10/43.69         &  45.12/44.78     &  46.3/45.4                &  36.0    &  44.03/43.65     &     49.3    &&      46.1(1.0)$^e$ &                 \\
 91 & 3p$^3$($^2$D)4p~$\mathrm{^3P_2}$ &   58.49/58.48         &  58.25/59.46     &  53.6/53.8                &  63.0    &  36.07/32.96     &             &&       53.5(4)$^i$         &                 \\    
 34 & 3p$^3$($^4$S)5p ~$\mathrm{^3P_2}$ &  179.06/173.11       &  179.14/178.85   &  184/186                  &  131.9   &  141.98/181.15   &             &&       188(13)$^e$  &                 \\    
                  &                 &                      &                  &                           &          &                  &             &&       185(15)$^i$  &                 \\    
 33 & 3p$^3$($^4$S)5p ~$\mathrm{^5P_3}$ &  156.05/149.48       &  155.78/153.94   &  158/159                  &  130.1   &  151.49/156.49   &             &&       615(50)$^i$         &                 \\   
 60 & 3p$^3$($^4$S)6p ~$\mathrm{^5P_3}$ &  352.19/329.68       &  351.23/346.94   &  352/350                  &  289.9   &  481.87/389.08   &             &&       265(20)$^i$         &                 \\   
 96 & 3p$^3$($^4$S)7p ~$\mathrm{^5P_3}$ &  596.11/621.01       &  601.79/673.21   &  631/624                  &  549.8   &                  &             &&       415(25)$^i$         &                 \\   
 23 & 3p$^3$($^4$S)3d ~$\mathrm{^3D_1}$ &  3.44/3.61           &  3.52/3.67       &  3.69/3.77                &  1.9     &  4.58/5.11       &     2.62    &&                &                 \\                        
 24 & 3p$^3$($^4$S)3d ~$\mathrm{^3D_2}$ &  3.44/3.60           &  3.52/3.66       &  3.68/3.77                &  1.9     &  4.45/4.97       &     2.61    &&                &    3.0(0.6)     \\
 25 & 3p$^3$($^4$S)3d ~$\mathrm{^3D_3}$ &  3.42/3.58           &  3.50/3.64       &  3.67/3.75                &  1.9     &  4.59/5.16       &     2.59    &&                &                 \\                        
 42 & 3p$^3$($^4$S)4d ~$\mathrm{^3D_1}$ &  11.87/12.36         &  11.97/12.48     &  12.1/12.6                &  6.0     &  6.27/7.09       &             &&       12.6(1.3)$^e$&                 \\  
 43 & 3p$^3$($^4$S)4d ~$\mathrm{^3D_2}$ &  11.85/12.34         &  11.96/12.46     &  12.1/12.6                &  6.0     &  6.49/7.43       &             &&       12.5(0.7)$^e$&    9.6(1.9)     \\
 44 & 3p$^3$($^4$S)4d ~$\mathrm{^3D_3}$ &  11.73/12.21         &  11.84/12.34     &  12.0/12.5                &  5.9     &  6.18/7.02       &             &&       12.9(1.1)$^e$&                 \\ 
 80 & 3p$^3$($^4$S)5d ~$\mathrm{^3D_1}$ &  37.18/38.84         &  38.38/39.73     &  36.0/37.8                &  10.1    &                  &             &&         41(3)$^e$  &                 \\
 78 & 3p$^3$($^4$S)5d ~$\mathrm{^3D_2}$ &  37.68/39.34         &  38.76/40.09     &  36.3/38.1                &  10.0    &                  &             &&                &    37(7.5)      \\
 79 & 3p$^3$($^4$S)5d ~$\mathrm{^3D_3}$ &  37.75/39.39         &  38.80/40.18     &  36.6/38.4                &  9.9     &                  &             &&         40(4)$^e$  &                 \\
 97 & 3p$^3$($^4$S)6d ~$\mathrm{^5D_4}$ &  266.27/269.21       &  264.35/251.99   &  165/163                  &  593.1   &                  &             &&         260(20)$^i$       &                 \\   
107 & 3p$^3$($^4$S)6d ~$\mathrm{^3D_1}$ &  134.65/135.55       &  141.41/141.03   &  108/115                  &  16.0    &                  &             &&                &                 \\   
106 & 3p$^3$($^4$S)6d ~$\mathrm{^3D_2}$ &  138.30/139.29       &  145.68/145.13   &  112/118                  &  15.9    &                  &             &&         75(7)$^b$ &    199(90)      \\
105 & 3p$^3$($^4$S)6d ~$\mathrm{^3D_3}$ &  142.01/143.06       &  150.12/149.33   &  116/123                  &  15.8    &                  &             &&                &                 \\   
\bottomrule
\end{tabular}
\tablefoot{B/C: Babushkin and Coulomb gauges. The values in the parentheses of the last two columns are the experimental errors.\\
\tablefoottext{a}{B-spline method by \cite{2006JPhB...39.2861Z}.}
\tablefoottext{b}{Relativistic Hartree–Fock calculation by \cite{1998ApJ...502.1010B}.}
\tablefoottext{c}{Configuration interaction calculation by \cite{2008ADNDT..94..561D}.}
\tablefoottext{d}{Multi-configurational Hartree-Fock calculation by \cite{FROESEFISCHER2006607}.}
\tablefoottext{e}{Laser spectroscopy by \cite{1997PhRvA..55.1836B} and \cite{RZerne_1997}.}
\tablefoottext{f}{Lifetime data excerpted from \cite{2006JPhB...39.2861Z} and original from \cite{1995ApJ...452..269F,Federman_1996}.}
\tablefoottext{g}{Beam-foil technique by \cite{1994ApJ...428..393B}.}
\tablefoottext{i}{High-frequency deflection technique by \cite{1990PhyS...42..540D}.}
\tablefoottext{j}{Electron-impact dissociation method by \cite{NJMason_1994}.}
}
\end{table*}

\subsection{Lifetimes}

Comparing the calculated lifetimes with the precise measured values can also help 
to evaluate the overall accuracy of the calculated transition data. 
Figure \ref{fig:tau} presents a comparison of our fine-tuned and ab initio lifetimes 
in Babushkin and Coulomb gauges, respectively (upper panel), as well as with the results from 
ZB2006 and DH2008 (lower panel). 
The ab initio and fine-tuned lifetimes agree very well with each other, except
for the levels with \#19, \#20, \#28, \#29, \#30, and \#58 in Table \ref{tab:energy}.
These levels also exhibit significant discrepancies with the results from ZB2006 and DH2008.
For levels with \#19, \#20, and \#58, the large discrepancies between ab initio and fine-tuned 
lifetimes are due to changes in the mixing coefficients of the CSFs caused by fine-tuning. 
For example, a level with \#58 of $\mathrm{3p^3(^4S)6p~^5P_1}$ has a percentage purity of 92\% 
in the ab initio calculation, which is close to pure $LS\!J$-coupling.
In contrast, fine-tuning reduced the percentage purity to 84\%, with increased mixing with 
$\mathrm{3p^{3}(^{2}D)4p~^{1}P_1}$ to 7\%, 
which shortens the lifetime of $\mathrm{3p^3(^4S)6p~^5P_1}$. 
The lifetimes of the triplet states $\mathrm{3s3p^{5}~^{3}P^{o}_{2,1,0}}$ (levels with \#28, \#29, and \#30) 
are partly determined by the TEOP transitions
$\mathrm{3s3p^{5}~^{3}P^{o}}$ - $\mathrm{3s^23p^{3}4p~^{3}P,~^5P}$ and the IC transitions
$\mathrm{3s3p^{5}~^{3}P^{o}}$ - $\mathrm{3s^23p^{4}~^{1}D,~^1S}$; these transitions are 
driven by electron correlation and are very challenging to compute accurately.
The triplet states are found to be
strongly mixed with $\mathrm{3p^{3}(^{2}D)nd~^{3}P^{o}(n=3-7)}$.
For these levels, the lifetimes in the Babushkin gauge are more stable, 
while fine-tuning causes larger changes in the Coulomb gauge.
This is consistent with the results discussed in Section \ref{ab-ft}, 
which namely show that fine-tuning has a greater impact on the transition data in
the Coulomb gauge.

Overall, the comparison in the lower panel of Figure \ref{fig:tau} shows that our results have better 
agreement with those from ZB2006 than with DH2008.
Among the 78 levels with three sets (ZB2006, DH2008 and our calculation) of lifetimes available, 
69 of them show agreement with the results from ZB2006 within the range of 
0.9 $\leq$  $\tau/\tau_\mathrm{ft}$  $\leq$ 1.1, while only 35 levels from DH2008 meet this criterion.
We also note that our computed lifetimes of the Rydberg states 
$\mathrm{3p^{3}(^4S)6d~^{5}D^{o}}$ (levels with No. \#97-\#101) quintet
and the $\mathrm{3p^{3}(^4S)7p~^{3}P}$ (levels with No. \#102-\#104) triplet also show 
significant differences from those of ZB2006. 
These levels also experience strong level mixing, especially the 
$\mathrm{3p^{3}(^4S)7p~^{3}P}$ triplet,
which has a percentage purity of only about 50\%.
For $\mathrm{3p^3(^4S)6d\,5D^o_{4}}$, our results, i.e. 266.27/269.21 ns from the
ab initio calculation and 264.35/251.99 ns from the
fine-tuned calculation, are in very good agreement with the experimental value 
of 260 $\pm$ 20 ns, which was obtained by \cite{1990PhyS...42..540D} using the high-frequency 
deflection technique (see Table \ref{tab:tau}). In contrast, the results from
ZB2006 are 165/163 ns, which are significantly lower than the experimental values. 

Table \ref{tab:tau} compares our calculated lifetimes with other theoretical calculations 
and experimentally measured results. 
Other calculation results include B-spline calculation of ZB2006, 
HFR calculation of \cite{1998ApJ...502.1010B}, 
CI calculation of DH2008, and MCHF calculation of \cite{FROESEFISCHER2006607}. 
In comparison to the others, the HFR results are generally much smaller, 
except for $\mathrm{3p3(^2P)4s\,^3P^o_{0,1,2}}$, $\mathrm{3p^3(^2D)4p\,^3P_{2}}$, 
and $\mathrm{3p^{3}(^{4}S)6d\,^5D^o_{4}}$. 
The theoretical results from the present calculations, as well as those from ZB2006 and DH2008,
are in very good agreement with the experimental 
measurements obtained by \cite{1997PhRvA..55.1836B} using laser spectroscopy and 
\cite{1994ApJ...428..393B} using the beam-foil technique; they are also in good 
agreement with the astronomical observations by \cite{1995ApJ...452..269F}, 
except for the $\mathrm{3p^3(^4S)4d\,3D_{1,2,3}}$ states, where the 
lifetimes from DH2008 are approximately half of the measured results by \cite{1997PhRvA..55.1836B}.
For the long-lived states, namely $\mathrm{3p^3(^4S)4s\,^5S^o_{2}}$, $\mathrm{3p^3(^4S)np\,^5P_{3} (n=5,6,7)}$, 
and $\mathrm{3p^3(^4S)6d\,^3D^o_{2}}$, 
there are significant differences between theoretical and experimental 
measurements, nor among different calculations and experimental results.
For these energy levels, which are dominated by weak transitions, 
both theoretical calculations and experimental measurements are challenging.
In addition, the measurements of lifetimes for long-lived states often have 
large uncertainties and further precise measurements are needed.

\section{Conclusion}\label{sec:conclusion}
The energy levels, transition data, and lifetimes of \si were calculated using both 
the ab initio and fine-tuning approaches based on the MCDHF and RCI methods. 
We found that fine-tuning provides more accurate mixing coefficients, 
which correctly position the energy levels.
It can also better improve the results in the Coulomb gauge, which is 
typically less accurate in ab initio calculations.
The corrections to the wave functions resulting from fine-tuning improve the consistency
of the transition data between the Babushkin and Coulomb gauges, thereby enhancing 
the accuracy of the transition data.

We conducted extensive comparisons with various theoretical and experimental results.
The accuracy level of the computed data was evaluated using a combination 
of gauge differences parameters and CFs. 
For applications requiring complete atomic datasets, we recommend using the 
fine-tuned data in the Babushkin gauge; for specific transitions, cross-validation 
with different sources is advised to ensure reliability. 
Notably, significant discrepancies were observed among different theoretical 
calculations for certain transitions, particularly for the solar-abundance 
diagnostic lines. The lack of precise experimental 
measurements complicates the assessment of the accuracy of different calculations. 
Further experimental efforts are essential to validating and constraining theoretical predictions.

The atomic data provided in this study can be used 
for astrophysical applications, 
not least non-LTE
stellar spectroscopic analyses
\citep[e.g.][]{2024ARep...68.1159K}.
In particular, the implications of these new data
on the solar sulphur abundance
were recently discussed in detail in \cite{2025A&A...703A..35A}.

\section*{Data availability}
The full versions of Table \ref{tab:energy} and Table \ref{tab:tr} are only available in electronic form at the CDS via 
anonymous ftp to cdsarc.u-strasbg.fr (130.79.128.5) or via http://cdsweb.u-strasbg.fr/cgi-bin/qcat?J/A+A/.

\begin{acknowledgements}
WL acknowledges the support from the National Key R\&D Program of China No. 2022YFF0503800, 
the National Natural Science Foundation of China (NSFC, Grant No. 12373058) and the specialized research fund
for State Key Laboratory of Solar Activity and Space Weather. 
AMA acknowledges support from the Swedish Research
Council (VR 2020-03940) the Crafoord Foundation via the Royal Swedish
Academy of Sciences (CR 2024-0015), and the European Union’s Horizon Europe
research and innovation programme under grant agreement No. 101079231
(EXOHOST). PJ acknowledges support from the Swedish Research Council (VR 2023-05367).
AMA and WL also acknowledge support from the 2024 Chinese Academy of Sciences
(CAS) President’s International Fellowship Initiative (PIFI).
WL thanks Yanting Li for the helpful discussion with the rfinetune code.
We would also like to thank the anonymous referee for providing useful 
comments that helped improve the original manuscript.
\end{acknowledgements}

\bibliographystyle{aa}
\bibliography{refs}

\begin{thebibliography}{82}
\expandafter\ifx\csname natexlab\endcsname\relax\def\natexlab#1{#1}\fi

\bibitem[{{Amarsi} {et~al.}(2025){Amarsi}, {Li}, {Grevesse}, \&
  {Jurewicz}}]{2025A&A...703A..35A}
{Amarsi}, A.~M., {Li}, W., {Grevesse}, N., \& {Jurewicz}, A.~J.~G. 2025, \aap,
  703, A35

\bibitem[{Anderson {et~al.}(2013)Anderson, Bergin, Maret, \&
  Wakelam}]{Anderson_2013}
Anderson, D.~E., Bergin, E.~A., Maret, S., \& Wakelam, V. 2013, The
  Astrophysical Journal, 779, 141

\bibitem[{{Beideck} {et~al.}(1994){Beideck}, {Schectman}, {Federman}, \&
  {Ellis}}]{1994ApJ...428..393B}
{Beideck}, D.~J., {Schectman}, R.~M., {Federman}, S.~R., \& {Ellis}, D.~G.
  1994, \apj, 428, 393

\bibitem[{{Berzinsh} {et~al.}(1997){Berzinsh}, {Caiyan}, {Zerne}, {Svanberg},
  \& {Bi{\'e}mont}}]{1997PhRvA..55.1836B}
{Berzinsh}, U., {Caiyan}, L., {Zerne}, R., {Svanberg}, S., \& {Bi{\'e}mont}, E.
  1997, \pra, 55, 1836

\bibitem[{{Bi{\'e}mont} {et~al.}(1998){Bi{\'e}mont}, {Garnir}, {Federman},
  {Li}, \& {Svanberg}}]{1998ApJ...502.1010B}
{Bi{\'e}mont}, E., {Garnir}, H.~P., {Federman}, S.~R., {Li}, Z.~S., \&
  {Svanberg}, S. 1998, \apj, 502, 1010

\bibitem[{{Bi{\'e}mont} {et~al.}(1996){Bi{\'e}mont}, {Storey}, \&
  {Zeippen}}]{1996A&A...309..991B}
{Bi{\'e}mont}, E., {Storey}, P.~J., \& {Zeippen}, C.~J. 1996, \aap, 309, 991

\bibitem[{{Bridges} \& {Wiese}(1967)}]{1967PhRv..159...31B}
{Bridges}, J.~M. \& {Wiese}, W.~L. 1967, Physical Review, 159, 31

\bibitem[{{Caffau} {et~al.}(2007){Caffau}, {Faraggiana}, {Bonifacio}, {Ludwig},
  \& {Steffen}}]{2007A&A...470..699C}
{Caffau}, E., {Faraggiana}, R., {Bonifacio}, P., {Ludwig}, H.-G., \& {Steffen},
  M. 2007, \aap, 470, 699

\bibitem[{{Carpenter} {et~al.}(1994){Carpenter}, {Robinson}, {Wahlgren},
  {Linsky}, \& {Brown}}]{1994ApJ...428..329C}
{Carpenter}, K.~G., {Robinson}, R.~D., {Wahlgren}, G.~M., {Linsky}, J.~L., \&
  {Brown}, A. 1994, \apj, 428, 329

\bibitem[{{Chen} {et~al.}(2001){Chen}, {Cheng}, \&
  {Johnson}}]{2001PhRvA..64d2507C}
{Chen}, M.~H., {Cheng}, K.~T., \& {Johnson}, W.~R. 2001, \pra, 64, 042507

\bibitem[{{Chen} \& {Msezane}(1997)}]{1997JPhB...30.3873C}
{Chen}, Z. \& {Msezane}, A.~Z. 1997, Journal of Physics B Atomic Molecular
  Physics, 30, 3873

\bibitem[{{Civi{\v{s}}} {et~al.}(2024){Civi{\v{s}}}, {Kramida}, {Zanozina},
  {Kubi{\v{s}}ta}, {Kubel{\'\i}k}, {Ferus}, \& {Chernov}}]{2024ApJS..274...32C}
{Civi{\v{s}}}, S., {Kramida}, A., {Zanozina}, E.~M., {et~al.} 2024, \apjs, 274,
  32

\bibitem[{Corrégé \& Hibbert(2004)}]{CORREGE200419}
Corrégé, G. \& Hibbert, A. 2004, Atomic Data and Nuclear Data Tables, 86, 19

\bibitem[{{Costa Silva} {et~al.}(2020){Costa Silva}, {Delgado Mena}, \&
  {Tsantaki}}]{2020A&A...634A.136C}
{Costa Silva}, A.~R., {Delgado Mena}, E., \& {Tsantaki}, M. 2020, \aap, 634,
  A136

\bibitem[{Cowan(1981)}]{Cowan_book}
Cowan, R.~D. 1981, The Theory of Atomic Structure and Spectra, 1st edn., Vol.~3
  (University of California Press)

\bibitem[{{da Silva} {et~al.}(2023){da Silva}, {D'Orazi}, {Palla}, {Bono},
  {Braga}, {Fabrizio}, {Lemasle}, {Spitoni}, {Matteucci}, {J{\"o}nsson},
  {Kovtyukh}, {Magrini}, {Bergemann}, {Dall'Ora}, {Ferraro}, {Fiorentino},
  {Fran{\c{c}}ois}, {Iannicola}, {Inno}, {Kudritzki}, {Matsunaga}, {Monelli},
  {Nonino}, {Sneden}, {Storm}, {Th{\'e}v{\'e}nin}, {Tsujimoto}, \&
  {Zocchi}}]{2023A&A...678A.195D}
{da Silva}, R., {D'Orazi}, V., {Palla}, M., {et~al.} 2023, \aap, 678, A195

\bibitem[{{Daflon} {et~al.}(2003){Daflon}, {Cunha}, {Smith}, \&
  {Butler}}]{2003A&A...399..525D}
{Daflon}, S., {Cunha}, K., {Smith}, V.~V., \& {Butler}, K. 2003, \aap, 399, 525

\bibitem[{{Deb} \& {Hibbert}(2006)}]{2006JPhB...39.4301D}
{Deb}, N.~C. \& {Hibbert}, A. 2006, Journal of Physics B Atomic Molecular
  Physics, 39, 4301

\bibitem[{{Deb} \& {Hibbert}(2008)}]{2008ADNDT..94..561D}
{Deb}, N.~C. \& {Hibbert}, A. 2008, Atomic Data and Nuclear Data Tables, 94,
  561

\bibitem[{{Delalic} {et~al.}(1990){Delalic}, {Erman}, \&
  {Kallne}}]{1990PhyS...42..540D}
{Delalic}, Z., {Erman}, P., \& {Kallne}, E. 1990, \physscr, 42, 540

\bibitem[{{Doering}(1990)}]{1990JGR....9521313D}
{Doering}, J.~P. 1990, \jgr, 95, 21313

\bibitem[{{Duffau} {et~al.}(2017){Duffau}, {Caffau}, {Sbordone}, {Bonifacio},
  {Andrievsky}, {Korotin}, {Babusiaux}, {Salvadori}, {Monaco},
  {Fran{\c{c}}ois}, {Sk{\'u}lad{\'o}ttir}, {Bragaglia}, {Donati}, {Spina},
  {Gallagher}, {Ludwig}, {Christlieb}, {Hansen}, {Mott}, {Steffen}, {Zaggia},
  {Blanco-Cuaresma}, {Calura}, {Friel}, {Jim{\'e}nez-Esteban}, {Koch},
  {Magrini}, {Pancino}, {Tang}, {Tautvai{\v{s}}ien{\.{e}}}, {Vallenari},
  {Hawkins}, {Gilmore}, {Randich}, {Feltzing}, {Bensby}, {Flaccomio},
  {Smiljanic}, {Bayo}, {Carraro}, {Casey}, {Costado}, {Damiani}, {Franciosini},
  {Hourihane}, {Jofr{\'e}}, {Lardo}, {Lewis}, {Morbidelli}, {Sousa}, \&
  {Worley}}]{2017A&A...604A.128D}
{Duffau}, S., {Caffau}, E., {Sbordone}, L., {et~al.} 2017, \aap, 604, A128

\bibitem[{{Durrance} {et~al.}(1983){Durrance}, {Feldman}, \&
  {Weaver}}]{1983ApJ...267L.125D}
{Durrance}, S.~T., {Feldman}, P.~D., \& {Weaver}, H.~A. 1983, \apjl, 267, L125

\bibitem[{Dyall {et~al.}(1989)Dyall, Grant, Johnson, Parpia, \& Plummer}]{EOL}
Dyall, K., Grant, I., Johnson, C., Parpia, F., \& Plummer, E. 1989, Comput.
  Phys. Commun., 55, 425

\bibitem[{{Ekman} {et~al.}(2014){Ekman}, {Godefroid}, \& {Hartman}}]{Ekman2014}
{Ekman}, J., {Godefroid}, M., \& {Hartman}, H. 2014, Atoms, 2, 215

\bibitem[{{Fawcett}(1986)}]{1986ADNDT..35..185F}
{Fawcett}, B.~C. 1986, Atomic Data and Nuclear Data Tables, 35, 185

\bibitem[{{Feaga} {et~al.}(2002){Feaga}, {McGrath}, \&
  {Feldman}}]{2002ApJ...570..439F}
{Feaga}, L.~M., {McGrath}, M.~A., \& {Feldman}, P.~D. 2002, \apj, 570, 439

\bibitem[{{Federman} \& {Cardelli}(1995)}]{1995ApJ...452..269F}
{Federman}, S.~R. \& {Cardelli}, J.~A. 1995, \apj, 452, 269

\bibitem[{Federman \& Cardelli(1996)}]{Federman_1996}
Federman, S.~R. \& Cardelli, J.~A. 1996, Physica Scripta, 1996, 158

\bibitem[{Fischer(1987)}]{Fischer_1987}
Fischer, C.~F. 1987, Journal of Physics B: Atomic and Molecular Physics, 20,
  4365

\bibitem[{{Fleming} \& {Hibbert}(1999)}]{1999PhST...83...44F}
{Fleming}, J. \& {Hibbert}, A. 1999, Physica Scripta Volume T, 83, 44

\bibitem[{{Froese Fischer}(2009)}]{Fischer2009}
{Froese Fischer}, C. 2009, Physica Scripta Volume T, 134, 014019

\bibitem[{{Froese Fischer} {et~al.}(2019){Froese Fischer}, Gaigalas,
  J{\"o}nsson, \& Biero\'n}]{Grasp2018}
{Froese Fischer}, C., Gaigalas, G., J{\"o}nsson, P., \& Biero\'n, J. 2019,
  Comput. Phys. Commun., 237, 184

\bibitem[{Froese~Fischer {et~al.}(2016)Froese~Fischer, Godefroid, Brage,
  J{\"o}nsson, \& Gaigalas}]{fischer.2016}
Froese~Fischer, C., Godefroid, M., Brage, T., J{\"o}nsson, P., \& Gaigalas, G.
  2016, Journal of Physics B: Atomic, Molecular and Optical Physics, 49, 182004

\bibitem[{{Froese Fischer} \& Tachiev(2004)}]{FROESEFISCHER20041}
{Froese Fischer}, C. \& Tachiev, G. 2004, Atomic Data and Nuclear Data Tables,
  87, 1

\bibitem[{{Froese Fischer} {et~al.}(2007){Froese Fischer}, Tachiev, Gaigalas,
  \& Godefroid}]{FROESEFISCHER2007559}
{Froese Fischer}, C., Tachiev, G., Gaigalas, G., \& Godefroid, M.~R. 2007,
  Computer Physics Communications, 176, 559

\bibitem[{{Froese Fischer} {et~al.}(2006){Froese Fischer}, Tachiev, \&
  Irimia}]{FROESEFISCHER2006607}
{Froese Fischer}, C., Tachiev, G., \& Irimia, A. 2006, Atomic Data and Nuclear
  Data Tables, 92, 607

\bibitem[{{Gaigalas} {et~al.}(2001){Gaigalas}, {Fritzsche}, \&
  {Grant}}]{2001CoPhC.139..263G}
{Gaigalas}, G., {Fritzsche}, S., \& {Grant}, I.~P. 2001, Computer Physics
  Communications, 139, 263

\bibitem[{{Gaigalas} {et~al.}(2017){Gaigalas}, {Froese Fischer}, {Rynkun}, \&
  {J{\"o}nsson}}]{2017Atoms...5....6G}
{Gaigalas}, G., {Froese Fischer}, C., {Rynkun}, P., \& {J{\"o}nsson}, P. 2017,
  Atoms, 5, 6

\bibitem[{{Gaigalas} {et~al.}(1997){Gaigalas}, {Rudzikas}, \& {Froese
  Fischer}}]{1997JPhB...30.3747G}
{Gaigalas}, G., {Rudzikas}, Z., \& {Froese Fischer}, C. 1997, Journal of
  Physics B Atomic Molecular Physics, 30, 3747

\bibitem[{{Gaigalas} {et~al.}(2024){Gaigalas}, {Rynkun}, {Domoto}, {Tanaka},
  {Kato}, \& {Kitovien{\.{e}}}}]{2024MNRAS.530.5220G}
{Gaigalas}, G., {Rynkun}, P., {Domoto}, N., {et~al.} 2024, \mnras, 530, 5220

\bibitem[{{Gaigalas} {et~al.}(2020){Gaigalas}, {Rynkun},
  {Rad{\v{z}}i{\={u}}t{\.{e}}}, {Kato}, {Tanaka}, \&
  {J{\"o}nsson}}]{2020ApJS..248...13G}
{Gaigalas}, G., {Rynkun}, P., {Rad{\v{z}}i{\={u}}t{\.{e}}}, L., {et~al.} 2020,
  \apjs, 248, 13

\bibitem[{{Gaigalas} {et~al.}(2003){Gaigalas}, {{\v{Z}}alandauskas}, \&
  {Rudzikas}}]{2003ADNDT..84...99G}
{Gaigalas}, G., {{\v{Z}}alandauskas}, T., \& {Rudzikas}, Z. 2003, Atomic Data
  and Nuclear Data Tables, 84, 99

\bibitem[{{Gaigalas} {et~al.}(2004){Gaigalas}, {Zalandauskas}, \&
  {Fritzsche}}]{2004CoPhC.157..239G}
{Gaigalas}, G., {Zalandauskas}, T., \& {Fritzsche}, S. 2004, Computer Physics
  Communications, 157, 239

\bibitem[{{Ganas}(1982)}]{1982PhLA...87..394G}
{Ganas}, P.~S. 1982, Physics Letters A, 87, 394

\bibitem[{{Grant}(1974)}]{1974JPhB....7.1458G}
{Grant}, I.~P. 1974, Journal of Physics B Atomic Molecular Physics, 7, 1458

\bibitem[{Grant(2007)}]{Grant2007}
Grant, I.~P. 2007, Relativistic Quantum Theory of Atoms and Molecules
  (Springer, New York)

\bibitem[{Hibbert(1974)}]{Hibbert_1974}
Hibbert, A. 1974, Journal of Physics B: Atomic and Molecular Physics, 7, 1417

\bibitem[{Hibbert(1975)}]{HIBBERT1975141}
Hibbert, A. 1975, Computer Physics Communications, 9, 141

\bibitem[{{Ho} \& {Henry}(1985)}]{1985ApJ...290..424H}
{Ho}, Y.~K. \& {Henry}, R.~J.~W. 1985, \apj, 290, 424

\bibitem[{{Judge}(1988)}]{1988MNRAS.231..419J}
{Judge}, P.~G. 1988, \mnras, 231, 419

\bibitem[{Jönsson {et~al.}(2023{\natexlab{a}})Jönsson, Gaigalas, Fischer,
  Bieroń, Grant, Brage, Ekman, Godefroid, Grumer, Li, \& Li}]{atoms11040068}
Jönsson, P., Gaigalas, G., Fischer, C.~F., {et~al.} 2023{\natexlab{a}}, Atoms,
  11

\bibitem[{Jönsson {et~al.}(2023{\natexlab{b}})Jönsson, Godefroid, Gaigalas,
  Ekman, Grumer, Li, Li , Brage, Grant, Bieroń, \& Fischer}]{atoms11010007}
Jönsson, P., Godefroid, M., Gaigalas, G., {et~al.} 2023{\natexlab{b}}, Atoms,
  11

\bibitem[{{Kamp} {et~al.}(2001){Kamp}, {Iliev}, {Paunzen}, {Pintado}, {Solano},
  \& {Barzova}}]{2001A&A...375..899K}
{Kamp}, I., {Iliev}, I.~K., {Paunzen}, E., {et~al.} 2001, \aap, 375, 899

\bibitem[{{Kaufman} \& {Martin}(1993)}]{1993JPCRD..22..279K}
{Kaufman}, V. \& {Martin}, W.~C. 1993, Journal of Physical and Chemical
  Reference Data, 22, 279

\bibitem[{{Korotin} \& {Kiselev}(2024)}]{2024ARep...68.1159K}
{Korotin}, S.~A. \& {Kiselev}, K.~O. 2024, Astronomy Reports, 68, 1159

\bibitem[{Kramida {et~al.}(2024)Kramida, {Yu.~Ralchenko}, Reader, \& {and NIST
  ASD Team}}]{NIST_ASD}
Kramida, A., {Yu.~Ralchenko}, Reader, J., \& {and NIST ASD Team}. 2024, {NIST
  Atomic Spectra Database (ver. 5.12), [Online]. Available:
  {\tt{https://physics.nist.gov/asd}} [2025, August 21]. National Institute of
  Standards and Technology, Gaithersburg, MD.}

\bibitem[{{Kurucz} \& {Peytremann}(1975)}]{1975SAOSR.362.....K}
{Kurucz}, R.~L. \& {Peytremann}, E. 1975, SAO Special Report

\bibitem[{{Laverick} {et~al.}(2019){Laverick}, {Lobel}, {Royer}, {Merle},
  {Martayan}, {van Hoof}, {Van der Swaelmen}, {David}, {Hensberge}, \&
  {Thienpont}}]{2019A&A...624A..60L}
{Laverick}, M., {Lobel}, A., {Royer}, P., {et~al.} 2019, \aap, 624, A60

\bibitem[{{Li} {et~al.}(2023{\natexlab{a}}){Li}, {Li}, {J{\"o}nsson}, {Amarsi},
  \& {Grumer}}]{2023ApJS..265...26L}
{Li}, M.~C., {Li}, W., {J{\"o}nsson}, P., {Amarsi}, A.~M., \& {Grumer}, J.
  2023{\natexlab{a}}, \apjs, 265, 26

\bibitem[{{Li} {et~al.}(2023{\natexlab{b}}){Li}, {J{\"o}nsson}, {Amarsi}, {Li},
  \& {Grumer}}]{2023A&A...674A..54L}
{Li}, W., {J{\"o}nsson}, P., {Amarsi}, A.~M., {Li}, M.~C., \& {Grumer}, J.
  2023{\natexlab{b}}, \aap, 674, A54

\bibitem[{Li {et~al.}(2023)Li, Gaigalas, Li, Chen, \& Jönsson}]{atoms11040070}
Li, Y., Gaigalas, G., Li, W., Chen, C., \& Jönsson, P. 2023, Atoms, 11

\bibitem[{Lundberg {et~al.}(2001)Lundberg, Li, \&
  J\"onsson}]{PhysRevA.63.032505}
Lundberg, H., Li, Z.~S., \& J\"onsson, P. 2001, Phys. Rev. A, 63, 032505

\bibitem[{Mason(1994)}]{NJMason_1994}
Mason, N.~J. 1994, Physica Scripta, 49, 578

\bibitem[{{M{\"u}ller}(1968)}]{1968ZNatA..23.1707M}
{M{\"u}ller}, D. 1968, Zeitschrift Naturforschung Teil A, 23, 1707

\bibitem[{{Nissen} {et~al.}(2007){Nissen}, {Akerman}, {Asplund}, {Fabbian},
  {Kerber}, {Kaufl}, \& {Pettini}}]{2007A&A...469..319N}
{Nissen}, P.~E., {Akerman}, C., {Asplund}, M., {et~al.} 2007, \aap, 469, 319

\bibitem[{Olsen {et~al.}(1988)Olsen, Roos, Jo/rgensen, \& Jensen}]{olsen.1988}
Olsen, J., Roos, B.~O., Jo/rgensen, P., \& Jensen, H. J.~A. 1988, The Journal
  of chemical physics, 89, 2185

\bibitem[{Papoulia {et~al.}(2019)Papoulia, Ekman, Gaigalas, Godefroid,
  Gustafsson, Hartman, Li, Radžiūtė, Rynkun, Schiffmann, Wang, \&
  Jönsson}]{Papoulia2019}
Papoulia, A., Ekman, J., Gaigalas, G., {et~al.} 2019, Atoms, 7

\bibitem[{{Papoulia} {et~al.}(2019){Papoulia}, {Ekman}, \&
  {J{\"o}nsson}}]{2019A&A...621A..16P}
{Papoulia}, A., {Ekman}, J., \& {J{\"o}nsson}, P. 2019, \aap, 621, A16

\bibitem[{{Podobedova} {et~al.}(2009){Podobedova}, {Kelleher}, \&
  {Wiese}}]{2009JPCRD..38..171P}
{Podobedova}, L.~I., {Kelleher}, D.~E., \& {Wiese}, W.~L. 2009, Journal of
  Physical and Chemical Reference Data, 38, 171

\bibitem[{{Savage} \& {Lawrence}(1966)}]{1966ApJ...146..940S}
{Savage}, B.~D. \& {Lawrence}, G.~M. 1966, \apj, 146, 940

\bibitem[{{Scott} {et~al.}(2015){Scott}, {Grevesse}, {Asplund}, {Sauval},
  {Lind}, {Takeda}, {Collet}, {Trampedach}, \& {Hayek}}]{2015A&A...573A..25S}
{Scott}, P., {Grevesse}, N., {Asplund}, M., {et~al.} 2015, \aap, 573, A25

\bibitem[{{Skillman} \& {Kennicutt}(1993)}]{1993ApJ...411..655S}
{Skillman}, E.~D. \& {Kennicutt}, Jr., R.~C. 1993, \apj, 411, 655

\bibitem[{{Sk{\'u}lad{\'o}ttir} {et~al.}(2015){Sk{\'u}lad{\'o}ttir},
  {Andrievsky}, {Tolstoy}, {Hill}, {Salvadori}, {Korotin}, \&
  {Pettini}}]{2015A&A...580A.129S}
{Sk{\'u}lad{\'o}ttir}, {\'A}., {Andrievsky}, S.~M., {Tolstoy}, E., {et~al.}
  2015, \aap, 580, A129

\bibitem[{Starace(1971)}]{PhysRevA.3.1242}
Starace, A.~F. 1971, Phys. Rev. A, 3, 1242

\bibitem[{Sturesson {et~al.}(2007)Sturesson, J{\"o}nsson, \&
  Froese~Fischer}]{sturesson.2007}
Sturesson, L., J{\"o}nsson, P., \& Froese~Fischer, C. 2007, Computer physics
  communications, 177, 539

\bibitem[{{Takeda} {et~al.}(2005){Takeda}, {Hashimoto}, {Taguchi}, {Yoshioka},
  {Takada-Hidai}, {Saito}, \& {Honda}}]{2005PASJ...57..751T}
{Takeda}, Y., {Hashimoto}, O., {Taguchi}, H., {et~al.} 2005, \pasj, 57, 751

\bibitem[{{Tayal}(1998)}]{1998ApJ...497..493T}
{Tayal}, S.~S. 1998, \apj, 497, 493

\bibitem[{{Tennyson}(2019)}]{2019asia.book.....T}
{Tennyson}, J. 2019, {Astronomical Spectroscopy. An Introduction to the Atomic
  and Molecular Physics of Astronomical Spectroscopy} (World Scientific)

\bibitem[{{Zatsarinny} \& {Bartschat}(2006)}]{2006JPhB...39.2861Z}
{Zatsarinny}, O. \& {Bartschat}, K. 2006, Journal of Physics B Atomic Molecular
  Physics, 39, 2861

\bibitem[{Zerne {et~al.}(1997)Zerne, Caiyan, Berzinsh, \&
  Svanberg}]{RZerne_1997}
Zerne, R., Caiyan, L., Berzinsh, U., \& Svanberg, S. 1997, Physica Scripta, 56,
  459

\bibitem[{{Zhang} {et~al.}(2013){Zhang}, {Palmeri}, {Quinet}, \&
  {Bi{\'e}mont}}]{2013A&A...551A}
{Zhang}, W., {Palmeri}, P., {Quinet}, P., \& {Bi{\'e}mont}, {\'E}. 2013, \aap,
  551, A136

\end{thebibliography}

\begin{appendix}
\onecolumn

\section{Energy levels and transition data.}

\begin{table*}[h]
\caption{\label{tab:energy} Energy levels and lifetimes from \ab and fine$-$tuning calculations.} 
\begin{tabular}{lll|rrr|cc|cc}
\hline
\midrule
&&&\multicolumn{3}{|c|}{E (cm$^{-1}$)}&\multicolumn{2}{c|}{\ab}&\multicolumn{2}{c}{Fine$-$tuned} \\
\cmidrule{4-6}\cmidrule{7-8}\cmidrule(l){9-10}
No. & Configuration & State  & \ab & fine$-$tuned & NIST ASD & $\tau_B$ (s) & $\tau_C$ (s)& $\tau_B$ (s) & $\tau_C$ (s)  \\     
\midrule
  1 &  $\mathrm{3p^4}$            &   $\mathrm{^3P_2}$   &        0.00  &        0.00  &       0.0000 &    & &  \\
  2 &  $\mathrm{3p^4}$            &   $\mathrm{^3P_1}$   &      388.63  &      395.37  &     396.0565 &    & &  \\
  3 &  $\mathrm{3p^4}$            &   $\mathrm{^3P_0}$   &      563.86  &      572.33  &     573.5957 &    & &  \\
  4 &  $\mathrm{3p^4}$            &   $\mathrm{^1D_2}$   &     9423.71  &     9258.76  &    9238.6090 &    & &  \\
  5 &  $\mathrm{3p^4}$            &   $\mathrm{^1S_0}$   &    22798.74  &    22276.12  &   22179.9542 &    & &  \\
  6 &  $\mathrm{3p^3(^4S)4s}$   &   $\mathrm{^5S^o_2}$ &    52206.08  &    52586.28  &   52623.6050 &     1.389E$-$05  &     1.444E$-$05  &    1.351E$-$05  &     1.404E$-$05 \\
  7 &  $\mathrm{3p^3(^4S)4s}$   &   $\mathrm{^3S^o_1}$ &    54980.58  &    55302.32  &   55330.7750 &     1.931E$-$09  &     1.964E$-$09  &    1.944E$-$09  &     1.985E$-$09 \\
  8 &  $\mathrm{3p^3(^4S)4p}$   &   $\mathrm{^5P_1}$   &    62952.47  &    63405.52  &   63446.0380 &     3.605E$-$08  &     3.567E$-$08  &    3.616E$-$08  &     3.606E$-$08 \\
  9 &  $\mathrm{3p^3(^4S)4p}$   &   $\mathrm{^5P_2}$   &    62963.20  &    63416.88  &   63457.1166 &     3.595E$-$08  &     3.558E$-$08  &    3.606E$-$08  &     3.596E$-$08 \\
 10 &  $\mathrm{3p^3(^4S)4p}$   &   $\mathrm{^5P_3}$   &    62980.48  &    63434.97  &   63475.0270 &     3.577E$-$08  &     3.540E$-$08  &    3.588E$-$08  &     3.579E$-$08 \\
 $-$ &  $-$ &  $-$ &  $-$ &  $-$ &  $-$ &  $-$ &  $-$ &  $-$ &  $-$  \\
\bottomrule
\end{tabular}
\tablefoot{Energy levels are given relative to the ground state and are compared with the NIST ASD data \citep{NIST_ASD} that originates from \cite{2024ApJS..274...32C}. The lifetimes are given in Babushkin ($\tau_B$) and Coulomb ($\tau_C$) gauges, respectively.
The excitation energies for No. 94 and 95 are not available in the NIST ASD. Considering the close degeneracies in these states, 
we used the energy value of No. 96 (3p$^3$($^4$S)7p $^5$P$_3$) as the energies for No. 94 and 95 in the fine-tuning procedure.
Full table is available at the CDS.}
\end{table*}

\begin{table*}[h]
\tiny
\caption{\label{tab:tr} Computed transition data.}
\begin{tabular}{ccc|cccccc|cccccc}
\hline
\midrule
&&& \multicolumn{6}{|c|}{\ab} & \multicolumn{6}{c}{\ft} \\
\cmidrule{4-9}\cmidrule(l){10-15}
&&& \multicolumn{2}{|c}{log($gf$)} & \multicolumn{2}{c}{$A$ (s$^{-1}$)} & \multicolumn{2}{c|}{Acc} & \multicolumn{2}{c}{log($gf$)} & \multicolumn{2}{c}{$A$ (s$^{-1}$)} & \multicolumn{2}{c}{Acc.} \\
\cmidrule(r){4-5}\cmidrule{6-7}\cmidrule(l){8-9}\cmidrule(l){10-11}\cmidrule(l){12-13}\cmidrule(l){14-15}
  Upper state                         &           Lower state       &   $\lambda~(\AA)$ &     B     &        C     &   B   &   C &     B   &   C  &    B      &     C      &     B   &   C &     B   &   C  \\
\midrule
$\mathrm{3p^3(^4S)6d~^3D^o_1}$  &   $\mathrm{3p^4~^3P_2}$ &       1247.11    &  $-$4.01   &     $-$3.98  &  1.39E+05  &     1.50E+05 &   C   &   C  &  $-$4.04   &   $-$4.01 &  1.29E+05   &    1.39E+05 &     C   &   C  \\
$\mathrm{3p^3(^4S)6d~^3D^o_2}$  &   $\mathrm{3p^4~^3P_2}$ &       1247.14    &  $-$2.86   &     $-$2.83  &  1.17E+06  &     1.26E+06 &   C   &   C  &  $-$2.90   &   $-$2.87 &  1.08E+06   &    1.17E+06 &     C   &   C  \\
$\mathrm{3p^3(^4S)6d~^3D^o_3}$  &   $\mathrm{3p^4~^3P_2}$ &       1247.16    &  $-$2.15   &     $-$2.12  &  4.32E+06  &     4.67E+06 &   C   &   C  &  $-$2.19   &   $-$2.16 &  3.93E+06   &    4.27E+06 &     C   &   C  \\
$\mathrm{3p^3(^4S)6d~^5D^o_3}$  &   $\mathrm{3p^4~^3P_2}$ &       1250.12    &  $-$6.43   &     $-$7.27  &  2.28E+02  &     3.25E+01 &   E   &   E  &  $-$5.82   &   $-$6.13 &  9.14E+02   &    4.54E+02 &     E   &   E  \\
$\mathrm{3p^3(^4S)6d~^5D^o_2}$  &   $\mathrm{3p^4~^3P_2}$ &       1250.12    &  $-$6.13   &     $-$6.43  &  6.37E+02  &     3.20E+02 &   E   &   E  &  $-$6.14   &   $-$6.45 &  6.20E+02   &    3.04E+02 &     E   &   E  \\
$\mathrm{3p^3(^4S)6d~^5D^o_1}$  &   $\mathrm{3p^4~^3P_2}$ &       1250.12    &  $-$6.50   &     $-$6.71  &  4.50E+02  &     2.79E+02 &   E   &   E  &  $-$6.56   &   $-$6.79 &  3.89E+02   &    2.28E+02 &     E   &   E  \\
$\mathrm{3p^3(^4S)6d~^3D^o_1}$  &   $\mathrm{3p^4~^3P_1}$ &       1253.30    &  $-$2.85   &     $-$2.82  &  1.99E+06  &     2.14E+06 &   C   &   C  &  $-$2.89   &   $-$2.86 &  1.84E+06   &    1.97E+06 &     C   &   C  \\
$\mathrm{3p^3(^4S)6d~^3D^o_2}$  &   $\mathrm{3p^4~^3P_1}$ &       1253.33    &  $-$2.40   &     $-$2.37  &  3.35E+06  &     3.60E+06 &   C   &   C  &  $-$2.44   &   $-$2.41 &  3.07E+06   &    3.30E+06 &     C   &   C  \\
$\mathrm{3p^3(^4S)6d~^5D^o_2}$  &   $\mathrm{3p^4~^3P_1}$ &       1256.34    &  $-$8.91   &     $-$7.70  &  1.04E+00  &     1.67E+01 &   E   &   E  &  $-$8.56   &   $-$7.86 &  2.31E+00   &    1.16E+01 &     E   &   E  \\
$\mathrm{3p^3(^4S)6d~^5D^o_1}$  &   $\mathrm{3p^4~^3P_1}$ &       1256.34    &  $-$6.48   &     $-$6.76  &  4.68E+02  &     2.44E+02 &   E   &   E  &  $-$6.66   &   $-$7.03 &  3.04E+02   &    1.31E+02 &     E   &   E  \\
 $-$ &  $-$ &  $-$ &  $-$ &  $-$ &  $-$ &  $-$ &  $-$ &  $-$ &  $-$ & $-$   &  $-$ &  $-$ &  $-$ & $-$\\
\bottomrule
\end{tabular}
\tablefoot{Only the first ten rows are shown. 
The wavelengths are vacuum wavelengths obtained based on the energy levels in the NIST ASD originated from \cite{2024ApJS..274...32C}. It should be noted that the wavelengths of the transitions associated with No. 94 and 95 (3p$^3$($^4$S)7p $^5$P$_1, 2$) were calculated using the energy value of No. 96 (3p$^3$($^4$S)7p $^5$P$_3$). Full table is available at the CDS.}
\end{table*}

\end{appendix}
\end{document}